\documentclass[12pt]{article}
\usepackage{epsfig,amsfonts,amssymb,amsmath,framed}
\usepackage{hyperref}
\usepackage{cite}
\input epsf.sty
\usepackage{verbatim,slashed}
\usepackage[margin=2.7cm]{geometry}

\newcommand{\m}{m}

\def\ZZZ{{\hbox{ Z\kern-1.6mm Z}}}
\def\RRR{{\hbox{ R\kern-2.4mm R}}}
\def\CCC{{\hbox{ C\kern-2.0mm C}}}
\def\zzz{{\hbox{z\kern-1mm z}}}

\newcommand{\f}{\frac}

\newcommand{\qeq}{{\hbox{=\kern-2.3mm ? \kern.5mm }}}
\renewcommand{\qeq}{=}

\newcommand{\ul}{\underline}

\newcommand{\non}{\nonumber}
\newcommand{\be}{\begin{eqnarray}}
\newcommand{\ee}{\end{eqnarray}}
\newcommand{\ben}{\begin{eqnarray}\displaystyle}
\newcommand{\een}{\end{eqnarray}}

\newcommand{\p}{\partial}
\newcommand{\sectiono}[1]{\section{#1}\setcounter{equation}{0}}

\def\one{{\hbox{ 1\kern-.8mm l}}}
\def\zero{{\hbox{ 0\kern-1.5mm 0}}}

\newcommand{\bea}[1]{\begin{eqnarray}\label{#1} }
\newcommand{\eea}{\end{eqnarray}}

\usepackage{bm}
\usepackage[table]{xcolor}
\def\mvnote#1{{ #1}}

\begin{document}

\begin{flushright}
HRI/ST/1801
\end{flushright}

\baselineskip 24pt

\begin{center}
{\Large \bf  Integrated Massive Vertex Operator in Pure Spinor Formalism}

\end{center}

\vskip .6cm
\medskip

\vspace*{4.0ex}

\baselineskip=18pt

\begin{center}

{\large 
\rm Subhroneel Chakrabarti$^{a}$, Sitender Pratap Kashyap$^{a}$, Mritunjay Verma$^{a,b}$ }

\end{center}

\vspace*{4.0ex}

\centerline{\it \small $^a$Harish-Chandra Research Institute, HBNI,
Chhatnag Road, Jhunsi,
Allahabad 211019, India}
\centerline{ \it \small $^b$International Centre for Theoretical Sciences,
 Hesarghatta,
Bengaluru 560089, India.}

\vspace*{1.0ex}
\centerline{\small E-mail: subhroneelchak,
sitenderpratap, mritunjayverma@hri.res.in}

\vspace*{5.0ex}

\centerline{\bf Abstract} \bigskip
We construct the integrated vertex operator for the first massive states of open superstrings with $(mass)^2=1/\alpha'$ in the pure spinor formalism of the superstring theory. This vertex operator is expressed in terms of the ten dimensional $\mathcal{N}=1$ superfields describing the massive supermultiplet which appear in the unintegrated vertex operator of the same states.

\vfill

%\begin{center}
%{\Huge Preliminary version}
%\end{center}

\vfill \eject

\baselineskip 18pt

\tableofcontents

\sectiono{Introduction } 
The pure spinor formalism is a  super-Poincar\'e covariant formalism \cite{Berkovits, Berkovits7, Berkovits3} (for review, see \cite{Berkovitsnewrev, Berkovits2, Berkovits12, Mafra1, Oliver1, Joost1}) of superstrings. This feature allows for an efficient way of computing the scattering amplitudes \cite{BerkovitsNekrasov, Berkovits6, Berkovits7, Berkovits5, Berkovits9, Berkovits_Mafra, Mafra_Gomez} making computations simpler. The equivalence between the pure spinor and the other superstring formalisms has been verified in many examples  \cite{Berkovits4, Berkovits8,Berkovits10}.

\vspace*{.06in}As in {the Ramond-Neveu-Schwarz} (RNS) formalism, the scattering amplitudes in the pure spinor formalism also involve computing worldsheet correlation functions of unintegrated as well as integrated vertex operators. However, unlike {the RNS formalism}, the gauge fixed worldsheet action in the pure spinor formalism does not arise from the gauge fixing of a reparametrization invariant action. Due to this, there is no elementary $b$ ghost in the pure spinor formalism. This makes the relation between the unintegrated and the integrated vertex operators in the pure spinor formalism less direct. So, even though the computation of amplitudes are easier to carry out in pure spinor formalism, the construction of the vertex operators (integrated as well as unintegrated) are considerably more involved as compared to the RNS formalism (see e.g., \cite{grassi4, Jusinskas1}). {In this paper, we propose an ansatz for the integrated vertex and explicitly show that it satisfies the relevant BRST condition demonstrating that it is the correct integrated vertex operator. We shall also give the arguments as to how to arrive at the ansatz. This paper gives an explicit construction of the integrated vertex for the first massive states in open superstrings. The same method can also be used for the construction of the massive integrated as well as unintegrated vertex operators in the pure spinor formalism.}

\vspace*{.06in}Restricting to {the} open strings for simplicity, the vertex operators in pure spinor formalism in ten dimensional flat spacetime are constructed in the super-Poincar\'e covariant manner using $\mathcal{N}=1$ superfields. In particular, to construct the unintegrated vertex operator $V$ for the states at mass level $n$, i.e. for $m^2=\f{n}{\alpha'}$, one first needs to construct ``basis elements'' with ghost number $1$ and conformal weight $n$ using the world sheet pure spinor fields. These basis elements\footnote{\label{footnote1}We shall refer to the products of worldsheet pure spinor variables which appear in the vertex operators, multiplied by some superfield, as basis elements. So, e.g., $\p\theta^\beta \lambda^\alpha$ in equation \eqref{vertex_ansatz} will be referred as basis element which multiplies the superfield $B_{\alpha\beta}$.} are then multiplied with an arbitrary 10 dimensional $\mathcal{N}=1$ superfield. The unintegrated vertex operator is the most general linear combination of such objects. The superfields appearing in this unintegrated vertex operator are fixed using the on-shell condition $QV=0$, where $Q$ is the BRST operator of the theory. The integrated vertex operator $U$ can then be determined using the relation $QU=\p_{\mathbb{R}} V$ {where the subscript $\mathbb{R}$ in the right hand side denotes the fact that the derivative is taken along the real axis.} 

\vspace*{.06in}For the massless states, both the unintegrated as well as the integrated vertex operators are explicitly known. This allows us to calculate tree as well as loop amplitudes involving massless states in the pure spinor formalism. In this paper, we shall focus on the first massive states. The open string spectrum at first massive level comprises of 128 bosonic and 128 fermionic degrees of freedom. These states form a massive spin 2 supermultiplet of 10 dimensional $\mathcal{N}=1$ supersymmetry. The 128 fermionic degrees of freedom are encoded in a spin $3/2$ field $\psi_{m\alpha}$. On the other hand, the $128$ bosonic degrees of freedom are encoded in a 3-form field $b_{mnp}$ carrying 84 degrees of freedom and a symmetric traceless field $g_{mn}$ carrying 44 degrees of freedom (see, e.g. \cite{witten}). 

\vspace*{.06in}To describe the first massive states in a super-Poincar\'e covariant manner, we introduce three basic superfields $\Psi_{m\alpha}, B_{mnp}$ and $G_{mn}$ whose theta independent components are $\psi_{m\alpha}, b_{mnp}$ and $g_{mn}$ respectively. The higher theta components of these superfields contain the same physical fields in a more involved manner \cite{theta_exp}. The unintegrated vertex operator describing these states was constructed in \cite{Berkovits1} and its theta expansion was done in \cite{theta_exp}. The superfields appearing in this vertex operator can be expressed in terms of the basic superfields $G_{mn}, B_{mnp}$ or $\Psi_{m\alpha}$. 

\vspace*{.06in}In this paper, our goal will be to construct the integrated form of the vertex operator for the first massive states. We shall use the defining relation $QU=\p_{\mathbb{R}} V$ for this purpose. As we shall see, the superfields appearing in $U$ can also be expressed in terms of the basic superfields $\Psi_{m\alpha}, B_{mnp}$ and $G_{mn}$.

\vspace*{.06in}Rest of the paper is organized as follows. In section \ref{sec:review}, we briefly review some of the elements of the pure spinor formalism and the first massive unintegrated vertex operator which are used in our
analysis. In section \ref{strategy}, we give our general strategy and the main results of this paper.  The equations \eqref{general_Ures} and \eqref{Group_theory_Ures} are our main equations which give the first massive integrated vertex operator in terms of the basic superfields $B_{mnp}, G_{mn}$ and $\Psi_{m\alpha}$. In section \ref{Construction4}, we give the details of our construction following the strategy given in section \ref{strategy}. Finally, we conclude with discussion in section \ref{discus}. While our ansatz once verified to be a solution does not require any further justification, we summarize the chain of reasoning in appendix \ref{motive} that led us to our proposed ansatz. Even though the solution does not depend on how we arrive at this ansatz, the arguments presented in the appendix \ref{motive} are nonetheless of value since they imply that one can replicate the same method quite readily for all higher massive states.

\color{black}
\section{Review of Some Pure Spinor Elements}
\label{sec:review}
In this section, we briefly recall some of the results of the minimal pure spinor formalism and the first massive states which will be needed in our analysis. We shall also describe some results regarding open strings which will be needed in this paper.

\subsection{Some pure spinor results}
We start by recalling some results about the open string world-sheet theory in the pure spinor formalism. We shall follow the conventions used in \cite{theta_exp}. {The open string world-sheet CFT in the pure spinor formalism in flat spacetime is described by the action
	\be
	S=\f{1}{\pi\alpha'}\int_{UHP} d^2z \left(\f{1}{2}\p X^m\bar\p X_m+p^{L}_\alpha\bar\p\theta^\alpha_L-w^L_\alpha\bar\p\lambda_L^\alpha +p^{R}_\alpha\p\theta^\alpha_R-w^R_\alpha\p\lambda_R^\alpha\right)\label{action_before}
	\ee
	where, $m=0,1,,\cdots,9$ and $\alpha=1,\cdots,16$. Further, we use the acronym $UHP$ and $LHP$ for upper and lower half of the complex plane. The $L$ and $R$ denote the left and right moving fields respectively on the world-sheet which will be related through the boundary conditions. All the worldsheet fields $X^{m},p^L_\alpha, w^L_\alpha,\theta_L^\alpha$ and $\lambda_L^\alpha$ and the corresponding right moving fields (with script $R$) are function of both $(z,\bar{z})$ off-shell. However, on making use of the equations of motion, namely
	\begin{gather}
	\bar\p \p X^m(z,\bar z)=0\non\\
	\bar{\p}\theta_L^\alpha(z,\bar{z})=0 \quad , \quad \p\theta_R^\alpha(z,\bar{z})=0 \quad, \quad \bar{\p}p^L_\alpha(z,\bar{z})=0 \quad , \quad \p p^R_\alpha(z,\bar{z})=0 \label{eom}\\
	\bar{\p}\lambda_L^\alpha(z,\bar{z})=0 \quad , \quad \p\lambda_R^\alpha(z,\bar{z})=0 \quad, \quad \bar{\p}w^L_\alpha(z,\bar{z})=0 \quad , \quad \p w^R_\alpha(z,\bar{z})=0,\non
	\end{gather}
	we find that the fields with subscript $L$ and $R$ become holomorphic and anti-holomorphic respectively. The $X^m$ fields satisfy the harmonic equation and hence it can be written as sum of holomorphic and anti-holomorphic fields. This means that $\p X^m$ and $\bar\p X^m$ are holomorphic and anti-holomorphic respectively. Besides the above equations of motion, we have to impose appropriate boundary conditions. These boundary conditions for the open strings are	
	\begin{gather} 
	\p X^m(z,\bar z)=\bar \p X^m(\bar{z},z)      \non\\
	\theta_L^\alpha(z,\bar{z})=\theta_R^\alpha(\bar{z},z)     \non\\
	\hspace{.9in}p^L_\alpha(z,\bar{z})=p^R_\alpha(\bar{z},z)  \;,\quad \textup{at}\; z=\bar{z} \label{bdy_cond}\\ 
	\lambda_L^\alpha(z,\bar{z})=\lambda_R^\alpha(\bar{z},z)     \non\\
	w^L_\alpha(z,\bar{z})=w^R_\alpha(\bar{z},z)     \non 
	\end{gather} 
Taking these boundary conditions into account and using the action \eqref{action_before}, we can derive various OPEs. The OPE between the various matter sector fields can be worked out to be
\be
%  X^m(z,\bar z)X^n(w,\bar w)  = -\f{\alpha'}{2}\eta^{mn}\biggl[\ln|z-w|^2 +\ln|z-\bar w|^2\biggl]+\cdots
% \non\\[.2cm]
 \p X^m(z,\bar z)\p X^n(w,\bar w) & =& -\f{\alpha' \eta^{mn}}{2{( z-w})^2}+\cdots
\non\\[.2cm]
 p^L_\alpha(z,\bar z)\theta_L^\beta(w,\bar w)&=&\f{\alpha'}{2}\f{\delta_\alpha^{\;\;\beta}}{z-w}+\cdots \\[.2cm]
   p^R_\alpha(\bar z,z)\theta_R^\beta(\bar w,w)&=&\f{\alpha'}{2}\f{\delta_\alpha^{\;\;\beta}}{\bar z-\bar w}+\cdots \non
%  p^R_\alpha(\bar z,z)\theta_L^\beta(w,\bar w)=\f{\alpha'}{2}\f{\delta_\alpha^{\;\;\beta}}{\bar z-w}+\cdots\quad,\quad  p^L_\alpha( z,\bar z)\theta_R^\beta(\bar w,w)=\f{\alpha'}{2}\f{\delta_\alpha^{\;\;\beta}}{ z-\bar w}+\cdots
\label{OPEbef}
\ee
It is cumbersome to work with both left and right moving fields and impose the boundary conditions each time. Fortunately, using the ``doubling trick'', we can combine the left and right moving fields into a single field. The left and right moving fields considered so far are defined only in the upper half plane with their values agreeing on the real axis. Using the doubling trick, we construct a field defined in the whole complex plane. Moreover, this requires only the boundary conditions and not the on-shell conditions following from the equations of motion. For example, the boundary condition \eqref{bdy_cond} allows us to combine $\theta_L^\alpha(z,\bar z)$ and $\theta_R^\alpha(\bar z,z)$ into a single field as
	\be 
	\theta^\alpha (z,\bar{z})\equiv
	\begin{cases}\theta_L^\alpha (z,\bar{z}) \quad \textup{for} \;\; z \;\;\in\; UHP \\ \theta_R^\alpha (\bar{z},z) \quad \textup{for} \;\; z \;\;\in\; LHP \label{thetadef}
	                               \end{cases}
	\ee
We can similarly define $p_\alpha,w_\alpha$ and $\lambda^\alpha$ in the whole complex plane. Furthermore, all of the holomorphically factorized quantities such as the vertex operators and the stress tensor can be defined in a similar manner. The $\theta^\alpha$ as defined in \eqref{thetadef} is holomorphic in the whole complex plane. It is instructive to see this explicitly. For this, we need to show that $\bar\p\theta^\alpha=0$ for $z \in \mathbb{C}$. For $z \in UHP$, we have
	\be
	\bar{\p} \theta^\alpha (z,\bar{z}) \vert_{UHP} = \bar{\p} \theta_L^\alpha (z,\bar{z}) = 0,
	\ee by virtue of equation of motion for $\theta_L^\alpha$. On the other hand, for $z\in LHP$, we have
	\be
	\bar{\p} \theta^\alpha (z,\bar{z}) \vert_{LHP} = {\bar\p} \theta_R^\alpha (\bar{z},z)  = 0
	\ee 
where, we have used the fact that the equation of motion for $\theta_R^\alpha(\bar z, z)$ in \eqref{eom} implies that it is independent of the first argument. This completes the proof that $\theta^\alpha(z)$ is indeed a holomorphic function in the whole complex plane. Identical proofs can also be given for other fields or their derivatives. Moreover, the OPEs involving $\theta^\alpha_{L,R}$ and $p_\alpha^{L,R}$ which follow from \eqref{OPEbef} can be combined into a single OPE as 
\be
p_\alpha(z)\theta^\beta(w)=\f{\alpha'}{2}\f{\delta_\alpha^{\;\;\beta}}{z-w}+\cdots
\ee
From now on, we shall work with the fields defined using the doubling trick. However, one can always go back to the expressions involving the original fields using equation \eqref{thetadef} and similar relations for other fields.} The worldsheet fields $p_\alpha, w_\alpha,\theta^\alpha$ and $\lambda^\alpha$ carry the conformal weights $1,1,0,0$ respectively. The field $\lambda^\alpha$ satisfies the pure spinor constraint
\be
\lambda^\alpha\gamma^{m}_{\alpha\beta}\lambda^\beta=0 \label{psconstr}
\ee
The $\gamma^{m}$ in above equation are the $16\times 16$ gamma matrices. The antisymmetrized product of these gamma matrices are referred as forms. So, e.g., $\gamma_{mnp}^{\alpha\beta}$ is called 3-form and so on.

\vspace*{.06in}The field $\lambda^\alpha$ and $w_\beta$ carry the ghost numbers $1$ and $-1$ respectively. All other worldsheet fields carry the $0$ ghost number. Due to the pure spinor constraint, the worldsheet field  $w_\alpha$ only appears in the following gauge invariant\footnote{Here by gauge invariance, we mean invariance under $w_{\alpha} \rightarrow w_{\alpha}+ \Lambda^{m}(\gamma_{m}\lambda)_{\alpha}$.} combinations
\be
N^{mn}=\f{1}{2}w_\alpha (\gamma^{mn})^\alpha_{\;\;\beta}\lambda^\beta\quad,\qquad
J=w_\alpha\lambda^\alpha
\ee
All the components of these variables are not independent. This fact is captured by the following non-trivial constraint between the currents $N_{mn}$ and $J$ \cite{Berkovits1}
\be
:N^{mn}\lambda^\alpha :(z)(\gamma_m)_{\alpha\beta}-\f{1}{2}:J\lambda^\alpha :(z)(\gamma^n)_{\alpha\beta}-\alpha'\gamma^n_{\alpha\beta}\p\lambda^\alpha(z)\ =\ 0
\label{identity}
\ee
Two other important supersymmetric invariant combinations of the theory are given by 
\be
d_\alpha&=&p_\alpha-\f{1}{2}\gamma^m_{\;\;\alpha\beta}\theta^\beta\partial X_m-\f{1}{8}\gamma^m_{\alpha\beta}\gamma_{m\sigma\delta}\theta^\beta\theta^{\sigma}\partial\theta^\delta
\non\\[.4cm]\Pi^m&=&\partial X^m+\f{1}{2}\gamma^m_{\alpha\beta}\theta^\alpha\partial\theta^\beta \label{susy_momenta}
\ee

The BRST operator of the theory is given in terms of $\lambda^\alpha$ and $d_\alpha$ to be\footnote{Having holomorphic fields defined 
	in the whole complex plane using doubling trick means that we can use the closed contour integrals $\oint$ in the usual manner even for the open string. }
\be
Q=\oint dz\ \lambda^\alpha(z) d_\alpha(z)
\ee
%which generates the following transformations
%\be
%\delta X^m=\lambda\gamma^m\theta\quad,\quad \delta\theta^\alpha=\lambda^\alpha\quad,\quad \delta \lambda^\alpha=0\quad,\quad \delta d_\alpha=-\Pi^m(\gamma_m\lambda)_\alpha\quad,\quad \delta w_\alpha=d_\alpha
%\ee
The OPE between various worldsheet operators is given by\footnote{Note the minus sign in front of the single pole in $N^{mn}N^{pq}$ OPE. There is a typo regarding this sign in \cite{Berkovits1}. We thank Nathan Berkovits for confirming this.}
\be
d_\alpha(z)d_\beta(w)=-\f{\alpha'\gamma^m_{\alpha\beta}}{2(z-w)}\Pi_m(w)+\cdots
\quad,\qquad
d_\alpha(z)\Pi^m(w)=\f{\alpha'\gamma^m_{\alpha\beta}}{2(z-w)}\partial\theta^\beta(w)+\cdots\non
\ee
\be
d_\alpha(z)V(w)=\f{\alpha'}{2(z-w)}D_\alpha V(w)+\cdots \quad,\qquad
\Pi^m(z)V(w)=-\f{\alpha'}{(z-w)}\partial^mV(w)+\cdots\non
\ee
\be
 \Pi^m(z)\Pi^n(w)=-\f{\alpha'\eta^{mn}}{2(z-w)^2}+\cdots\quad,\qquad N^{mn}(z)\lambda^\alpha(w)=\f{\alpha'(\gamma^{mn})^\alpha_{\;\;\beta}}{4(z-w)}\ \lambda^\beta(w)+\cdots
\non
\ee
\be
J(z)J(w)=-\f{(\alpha')^2}{(z-w)^2}+\cdots\quad,\qquad  J(z)\lambda^\alpha(w)=\f{\alpha'}{2(z-w)}\lambda^\alpha(w)+\cdots\non
\ee
\be
N^{mn}(z)N^{pq}(w)=-\f{3(\alpha')^2}{2(z-w)^2}\eta^{m[q}\eta^{p]n}-\f{\alpha'}{(z-w)}\Bigl(\eta^{p[n}N^{m]q}-\eta^{q[n}N^{m]p}\Bigl)+\cdots\label{OPE_eq}
\ee
In the above OPEs, $\p_m$ is the derivative with respect to the spacetime coordinate $X^m$, $\p$ is the derivative with respect to the world-sheet coordinate, $V$ denotes an arbitrary superfield and $D_\alpha$ is the supercovariant derivative given by
 \be D_\alpha\equiv\p_\alpha+\gamma^m_{\alpha\beta}\theta^\beta\p_m\label{susy_derivative}
\ee

This supercovariant derivative satisfies the identity
\be
\lbrace D_\alpha, D_\beta\rbrace= 2 (\gamma^m)_{\alpha \beta} \partial_m \quad\implies\qquad (\gamma_m)^{\alpha\beta}D_\alpha D_\beta = \f{1}{16}\p_m
\ee

The matter and the ghost stress energy tensors of the theory are given by 
\be
T_m = -\f{1}{\alpha'}:\Pi^m \Pi_m: \ -\ \f{2}{\alpha'}:d_\alpha \p\theta^\alpha: \qquad,\quad T_g =\f{2}{\alpha'}w_\alpha \p\lambda^\alpha\label{stress_tensor}
\ee

The total stress tensor $T$ is given by the sum of $T_m$ and $T_g$\footnote{It is possible to express the ghost stress tensor $T_g$ in terms of the currents $N^{mn}$ and $J$ (see, e.g., \cite{Berkovits2}). However, we shall not need this expression. For our purposes, equation \eqref{TNOPE} is sufficient.}. The Lorentz current $N^{mn}$ is a primary operator with respect to the stress energy tensor. This follows due to the OPE
\be
T_g(z)N^{mn}(w)= \f{N^{mn}(w)}{(z-w)^2}+\f{\p N^{mn}(w)}{z-w}+\cdots\label{TNOPE}
\ee
and the fact that the matter and the ghost sector fields do not have any non trivial OPE between them. 

%\subsection{Some results regarding the first massive states}
\vspace*{.07in}After briefly reviewing the basics, we now turn to the first massive unintegrated vertex operator \cite{Berkovits1, theta_exp}. There are 128 fermionic and 128 bosonic degrees of freedom at the first massive level of the open string spectrum. The fermionic degrees of freedom are contained in a spin-$3/2$ field $\psi_{m\alpha}$ whereas the bosonic degrees of freedom are contained in a traceless symmetric tensor $g_{mn}$ and a 3-form field $b_{mnp}$. These fields are demanded to satisfy
\be 
\p^m\psi_{m\alpha}=0 \quad ; \quad \gamma^{m\alpha\beta} \psi_{m\beta}=0 \quad ;\quad \p^m b_{mnp}=0 \quad ;\quad
\eta^{mn}g_{mn}=0\quad ;\quad  \p^m g_{mn}=0   \label{constraints_theta_ind_comp}
\ee
These constraints ensure that the number of independent components in the fields $\psi_{m\beta}, b_{mnp}$ and $g_{mn}$ are $128, 84$ and $44$ respectively. These fields form a massive spin-2 supermultiplet in 10 dimensions. To describe the system in a supersymmetric invariant manner, we introduce basic superfields $\Psi_{m\alpha}, B_{mnp}$ and $G_{mn}$ whose theta independent components are $\psi_{m\alpha}, b_{mnp}$ and $g_{mn}$ respectively. The higher components of these basic superfields contain the same physical fields in a more involved manner.

\vspace*{.06in} At the first mass level, the unintegrated vertex operator of the open string is given by \cite{Berkovits1}
\ben
V\ =\ :\partial \theta^\beta \lambda^\alpha B_{\alpha\beta}: 
+:d_\beta\lambda^\alpha C^\beta_{\;\alpha}:+:\Pi^m\lambda^\alpha H_{m\alpha}:
+:N^{mn}\lambda^\alpha F_{\alpha mn}:                                                    \label{vertex_ansatz}
\een
where, the superfields appearing in the above expression are given in terms of the basic superfields $B_{mnp}$ and $\Psi_{m\alpha}$ to be \cite{Berkovits1}
\be 
&&H_{s\alpha}=\f{3}{7}(\gamma^{mn})_{\alpha}^{\;\;\beta}D_{\beta}B_{mns}=-72 \Psi_{s\alpha}\;, \quad C_{mnpq}=\f{1}{2}\p_{[m}B_{npq]} \;,  \non\\[.2in]
&&\hspace{.2 in}F_{\alpha mn} =\f{1}{8}\biggl(7\p_{[m}H_{n]\alpha} + \p^q(\gamma_{q[m})_\alpha^{\;\;\beta}H_{n]\beta}\biggl)
\label{2.14}
\ee
The normal ordering $:\;:$ is defined as 
\be 
:AB:(z) \equiv \frac{1}{2 \pi i} \oint_z  \frac{dw}{w-z} A(w)B(z)  
\ee
 where, $A$ and $B$ are any two operators and the contour surrounds the point $z$.

\vspace*{.06in}The basic superfields at the first massive level, namely, $B_{mnp}, \Psi_{m\alpha}$ and $G_{mn}$ satisfy the superspace equations\footnote{To go from position to momentum space and vice versa, we use the convention $\;\p_m\rightarrow ik_m\;\;\mbox{and}\;\; k_m\rightarrow -i\p_m$. We shall do calculations mostly in the momentum space but express the final result in the position space using this rule.} \cite{theta_exp}
 \be
D_{\alpha}G_{sm} = 16\ \p^{p}(\gamma_{p(s}\Psi_{m)})_\alpha \label{D_Gmn1}
\ee
\be
D_\alpha B_{mnp}=12 (\gamma_{[mn}\Psi_{p]})_\alpha - 24\alpha' \p^t \p_{[m}(\gamma_{|t|n}\Psi_{p]})_\alpha                             \label{D_Bmnp_general}
\ee
\be
D_{\alpha}\Psi_{s\beta}= \f{1}{16} G_{sm}\gamma^{m}_{\alpha\beta}+\f{1}{24}\p_mB_{nps}(\gamma^{mnp})_{\alpha\beta}
-\f{1}{144}\p^mB^{npq}(\gamma_{smnpq})_{\alpha\beta} \label{D_Psi1}
\ee
and the constraints
\be
 (\gamma^{m})^{\alpha \beta}\Psi_{m\beta}=0\quad ; \quad \p^m \Psi_{m\beta}=0\quad ; \quad  \p^m B_{mnp}=0 \quad ; \quad  \p^m G_{m n}=0 \;\; \;; \;\; \eta^{mn}G_{mn}=0\label{cons_theta=0}
\ee

\subsection{Some results regarding open strings}
\label{sec:convention}

For the open strings, the vertex operators live on the boundary, i.e., on the real axis in the complex plane. This means that in the BRST equation {$QU=\p_{\mathbb{R}} V$, the derivative in the right hand side is along the real axis (represented by the subscript $\mathbb{R}$. For comparing the left and right hand side of this equation, we shall need to express the partial derivative in the right hand side to derivative with respect to the world-sheet fields $X^m$ and $\theta^\alpha$. For this, we first convert the derivative along the real axis into the holomorphic derivative as follows. If $x$ denotes the coordinate along the real axis, {then the derivative of an arbitrary function $f$ along the real axis can be written as\footnote{We define $z = x+iy$ and $\bar{z}= x-iy$ with $x,y \in \mathbb{R}$.}}
\be
 \p_{\mathbb{R}}f = \f{\p f}{\p x}\ =\ \left(\f{\p z}{\p x}\f{\p f}{\p z}\ +\ \f{\p \bar z}{\p x}\f{\p f}{\p \bar z}\right)\Bigl|_{\bar z= z}\ =\ \left(\f{\p f}{\p z}\ +\ \f{\p f}{\p \bar z}\right)\Bigl|_{\bar z= z}\label{del_der}
\ee
Now, for the open strings, the left and right moving fields living on the world-sheet are identified along the real axis as in \eqref{bdy_cond}. Thus, any function along the real axis (such as the vertex operator) can be expressed only in terms of either left moving or right moving fields (or the fields defined using the doubling trick as in \eqref{thetadef}). Working with the left moving fields, we can use the chain rule to write
\be
\f{\p f}{\p z} \ =\ \f{\p f}{\p X^m}\f{\p X^m}{\p z} + \f{\p f}{\p \theta_L^\alpha}\f{\p \theta_L^\alpha}{\p z}
\quad,\qquad
\f{\p f}{\p \bar z} \ =\ \f{\p f}{\p X^m}\f{\p X^m}{\p \bar z} + \f{\p f}{\p \theta_L^\alpha}\f{\p \theta_L^\alpha}{\p \bar z}
\ee
Using the equation of motion for $p^L_\alpha$, namely, $\bar\p\theta_L^\alpha=0$ and the above equations, we obtain
\be
\p_\mathbb{R}f &=& \f{\p f}{\p X^m}\f{\p X^m}{\p z}\ +\ \f{\p f}{\p X^m}\f{\p X^m}{\p \bar z}\Bigl|_{\bar z= z}\ +\ \ \f{\p f}{\p \theta_L^\alpha}\f{\p \theta_L^\alpha}{\p z} \non\\[.4cm]
&=&2\f{\p f}{\p X^m}\f{\p X^m}{\p z}\ +\ \f{\p f}{\p \theta_L^\alpha}\f{\p \theta_L^\alpha}{\p z}
\ee
Now, for the left moving fields, the SUSY momenta and the supercovariant derivatives are given by
\be
\Pi^m_L&=&\partial X^m+\f{1}{2}\gamma^m_{\alpha\beta}\theta_L^\alpha\partial\theta_L^\beta\quad,\qquad D_\alpha^L = \f{\p}{\p\theta^\alpha_L}+\gamma^m_{\alpha\beta}\theta^\beta_L\p_m\label{acsd}
\ee
Using these, we obtain
\be
\p_\mathbb{R}f  
&=&2\Pi^m_L\p_m f + \p\theta_L^\alpha D^L_\alpha f\label{factor_2b}
\ee
If we had worked with the right moving fields, instead of the above equation, we would have obtained
\be
\p_\mathbb{R}f  
&=&2\Pi^m_R\p_m f + \bar\p\theta_R^\alpha D^R_\alpha f\label{factor_2a}
\ee
where, $\Pi^m_R$ and $D_\alpha^R$ are given by definitions similar to \eqref{acsd} but with $\theta^\alpha_L$ and $\p$ replaced by $\theta^\alpha_R$ and $\bar\p$ respectively. 

\vspace*{.07in}Since we are on the real axis, we can replace the left moving variables of \eqref{factor_2b} or the right moving variables of \eqref{factor_2a} in terms of fields defined on the whole complex plane using the doubling trick. Doing this, we obtain
\be
\p_\mathbb{R}f  
&=&2\Pi^m\p_m f + \p\theta^\alpha D_\alpha f\label{factor_2}
\ee
where $\Pi^m$ and $D_\alpha$ are given in \eqref{susy_momenta} and \eqref{susy_derivative} respectively.

\vspace*{.07in}We shall make use of the identity \eqref{factor_2} while computing the right hand side of the BRST equation $QU=\p_{\mathbb{R}} V$. Moreover, throughout the draft, the world-sheet derivatives $\p$ will denote the holomorphic derivative. In the places where it is derivative along the real axis (e.g., the right hand side of $QU=\p_{\mathbb{R}} V$), it can be easily converted to the holomorphic derivative using the identity \eqref{del_der}.}

\color{black}

\section{General Strategy and The Main Result} \label{strategy}
The integrated massive vertex $U$ is constructed following a series of steps which can be summarized quite succinctly. In this section, we give the general strategy as a series of steps while the subsequent section will provide the details of these steps. First let us state our goal clearly. All vertex operators (integrated or unintegrated) are schematically of the form $ \hat{O} A$, where $\hat{O}$ is a worldsheet operator of appropriate conformal weight and ghost number constructed out of ($\Pi^m, d_\alpha, \p \theta^\alpha, N^{mn}, J, \lambda^\alpha$) and their worldsheet derivatives and $A$ is a superfield whose tensor-spinor structure is such that $\hat{O}A$ is Lorentz invariant. As mentioned in footnote \ref{footnote1}, the operators $\hat O$ will be referred as basis elements. We know the expression of the unintegrated vertex \eqref{vertex_ansatz} in terms of the superfields which describe the massive supermultiplet (i.e. any one of $\Psi_{s \alpha}, B_{mnp}$ or $G_{mn}$). Our goal is to find $U$ in terms of the same superfields describing the massive multiplet such that it satisfies $QU = \p_{\mathbb{R}} V$.

\vspace*{.06in}The steps for the construction of the first massive vertex operator $U$ are as follows :
\begin{itemize}
	\item \textbf{Step 1 :} Write all possible worldsheet operators with conformal weight 2 and ghost number zero using $\Pi^m, d_\alpha, \p \theta^\alpha, N^{mn}, J, \lambda^\alpha$, noting that worldsheet derivative (denoted by $\p$) can increase the weight of any operator on worldsheet by 1. Contract each of these operators by an arbitrary superfield with appropriate index structure to obtain a Lorentz invariant combination. The most general $U$ is the sum of all these possible terms. 
	
	\item \textbf{Step 2 :} Compute $QU$ using the OPEs given in \eqref{OPE_eq}. Also compute the worldsheet derivative $\p_{\mathbb{R}} V$ of the unintegrated vertex operator.
	
	\item \textbf{Step 3 :} The pure spinor constraint \eqref{psconstr} and the OPEs \eqref{OPE_eq} imply several non trivial identities relating a specific subset of the basis operators of a given conformal weight and ghost number. List all such identities and express them in the form $I=0$. %The identities following from the pure spinor constraint can be obtained using \eqref{identity} by ``multiplying'' with object of appropriate conformal weight. Such identities can be expressed in the form $I=0$.
	
	\item \textbf{Step 4 :} To take into account the constraint identities, introduce Lagrange multipliers and set up the equation $QU -\p_{\mathbb{R}} V -IK =0$ (where $K$ denotes the Lagrange multiplier). The inclusion of $I$ ensures that all operator basis constructed in step 1 now can be treated as linearly independent. Instead of introducing the Lagrange multipliers, one can also directly eliminate some basis operators in favor of others.
	
	\item \textbf{Step 5 :} Express each of the arbitrary superfields in $U$ as a generic linear combination of $\Psi_{m \alpha}, B_{mnp}$ and $G_{mn}$ and their space time derivatives. The correct number of terms in each ansatz can be determined by using the representation theory of SO(9) which is the little group for the massive states in 10 dimensions. The number of times $\Psi_{m \alpha}, B_{mnp}$ and $G_{mn}$ will appear in a given ansatz is same as the number of \textbf{128}, \textbf{84} and \textbf{44} representations of SO(9) respectively in the superfield. This can be figured out by analyzing the index structure of the superfield in the rest frame.% and the methodology of covariantizing the rest frame results developed in (cite our paper).
	
	\item \textbf{Step 6 :} Substitute the ansatz of step 5 in the equations obtained in step 4. These lead to a set of linear algebraic equations for the unknown co-efficients appearing in the ansatz.
	
	\item \textbf{Step 7 :} Solve these linear equations. Plugging the solutions back allows us to express $U$ completely in terms of the superfields that describe the massive supermultiplet.
	
\end{itemize}
Following this procedure, the final form of the first massive integrated vertex operator is obtained to be
\be
U&=&\ :\Pi^m\Pi^n F_{mn}:\;+\;:\Pi^m d_\alpha F_m^{\;\;\alpha}:\;+\;:\Pi^m\partial\theta^\alpha G_{m\alpha}:
\;+\;:\Pi^m N^{pq}F_{mpq}:\nonumber\\
&&+\;\;:d_\alpha d_\beta K^{\alpha\beta}:\;+\;:d_\alpha\partial\theta^\beta F^\alpha_{\;\;\beta}:\;+\;:d_\alpha N^{mn}G^\alpha_{\;\;mn}:\;+\;:\partial\theta^\alpha\partial\theta^\beta H_{\alpha\beta}:
\nonumber\\
&&+\;:\partial\theta^\alpha N^{mn} H_{mn\alpha}:\;+\;\;:N^{mn}N^{pq}G_{mnpq}:
\label{general_Ures}
\ee
where, the superfields appearing in \eqref{general_Ures} are given in position space by
\be
F_{mn}&=&-\f{18}{\alpha'}G_{mn}\quad\;,\qquad F_m^{\;\alpha}\ =\ \f{288}{\alpha'} (\gamma^r)^{\alpha\beta}\p_r\Psi_{m\beta}
 \quad\;\;, \qquad G_{m\alpha}\ =\ -\f{432}{\alpha'}\Psi_{m\alpha} \non\\[.5cm]
 F_{mpq}&=&\f{12}{(\alpha')^2}B_{mpq}-\f{36}{\alpha'}\p_{[p}G_{q]m}\quad,\quad\;\; K^{\alpha \beta}\ =\ -\f{1}{(\alpha')^2}\; \gamma_{mnp}^{\alpha \beta}B^{mnp}  \non\\[.5cm]
F^{\alpha}_{\;\;\;\beta}&=&-\f{4}{\alpha'} (\gamma^{mnpq})^{\alpha}_{\;\;\;\beta}\p_mB_{npq}\qquad,\quad\;\;
 G^{ \alpha}_{mn}\ =\ \f{48}{(\alpha')^2}\gamma_{[m}^{ \alpha\sigma}\Psi_{n]\sigma}+\f{192}{\alpha'} \gamma_r^{ \alpha\sigma}\p^r\p_{[m}\Psi_{n]\sigma}
\non\\[.5cm]
H_{\alpha \beta}\ &=&\ \f{2}{\alpha'}\gamma^{mnp}_{\alpha \beta}B_{mnp}\quad\qquad\qquad\;\;,\quad H_{mn\alpha}\ =\ -\f{576}{\alpha'}\; \p_{[m}\Psi_{n]\alpha}-\f{144}{\alpha'}\p^q(\gamma_{q[m})_\alpha^{\;\;\sigma}\Psi_{n]\sigma}\non\\[.5cm]
G_{mnpq}&=&\f{4}{(\alpha')^2}\p_{[m}B_{n]pq}+\f{4}{(\alpha')^2}\p_{[p}B_{q]mn}-\f{12}{\alpha'}\p_{[p}\p_{[m}G_{n]q]}\label{Group_theory_Ures}
\ee
It can be explicitly verified that the integrated vertex operator constructed here is a primary operator with respect to the stress energy tensor of the theory\footnote{We thank Nathan Berkovits for raising this issue.}. The 3rd and the 4th order poles of the OPE between the total stress tensor $T$ and the vertex operator $U$ given in \eqref{general_Ures} vanish identically for the solution given in \eqref{Group_theory_Ures} on using the conditions \eqref{cons_theta=0}. The full computation, on using the expression of the matter stress tensor given in \eqref{stress_tensor} and the OPE between $T_g$ and $N^{mn}$ given in \eqref{TNOPE}, gives
\be
T(z)U(w) = \f{2U(w)}{(z-w)^2}+\f{\p U(w)}{z-w}+\cdots\label{TUOPE}
\ee 
which confirms that the integrated vertex operator $U$ is a world-sheet primary operator of conformal weight 2 with respect to the stress energy tensor.

\color{black}
\section{Details of the Derivation}
\label{Construction4}
In this section, we give the details of the procedure outlined in the previous section. 
To construct the integrated vertex operator for the massive states, we start by noting that the relation between the integrated and unintegrated vertex operator 
is given by
\be
QU(z)=\p_{\mathbb{R}} V(z)\qquad\implies \qquad\f{1}{2\pi i}\oint_z\ dw\ \lambda^\alpha(w) d_\alpha(w) U(z)=\p_{\mathbb{R}}V(z) 
\label{massive_for}
\ee
We shall derive the integrated vertex by first writing down the most general form of the integrated vertex in terms of arbitrary superfields and then use the above equation to determine these superfields. 
\subsection{Ingredients of Equation of Motion}
As mentioned earlier, the integrated vertex operator describing the physical states at mass level $n$, i.e., $m^{2}=\f{n}{\alpha'}$ is constructed out of 
objects with ghost number $0$ and conformal dimension $n+1$. These Lorentz and SUSY invariant objects are constructed using the pure spinor variables 
$\Pi^m, \p\theta^\alpha,d_\alpha, \lambda^\alpha, J \textup{ and } N^{mn}$. Moreover, as argued in appendix \ref{polynomial_dependence}, we can choose
the integrated vertex to be independnet of the $\bar\lambda\lambda$ factors. Consequently, the most general 
integrated vertex operator at first massive level ($n=1$) of the open string can be written as\footnote{Inside a normal ordering, the order of the
operators matters if they have non trivial OPE between them (see e.g., chapter 6 of \cite{Fransicso} ).
Hence, for comparing various expressions (e.g., LHS and RHS of $QU=\p_{\mathbb{R}} V$), we need to have the same ordering of the world-sheet operators inside normal ordering. 
However, during the intermediate stages of the calculation, the operators may not occur in the same order and we need to bring them in a given fixed order.
We shall use the following convention for the ordering of the world-sheet operators from left to right if more than one of them appear inside normal ordering
: $\Pi^m, d_\alpha,\p\theta^\alpha,  N^{mn}, J, \lambda^\alpha$. If the operators in some terms are not in this order, we shall bring them in this order using OPEs. 
An example of this is given in equation \eqref{example_ref}.}
\be
U&=&:\partial^2\theta^\alpha C_\alpha:\;+\;:\partial\Pi^mC_m:\;+\;:\partial d_\alpha E^\alpha:\;+\;:(\partial J)C:\;+\;:\partial N^{mn}C_{mn}:
\nonumber\\
&&+\;\;:\Pi^m\Pi^n F_{mn}:\;+\;:\Pi^m d_\alpha F_m^{\;\;\alpha}:\;+\;:\Pi^m N^{pq}F_{mpq}:\;+\;:\Pi^mJF_m:\;+\;:\Pi^m\partial\theta^\alpha G_{m\alpha}:
\nonumber\\
&&+\;\;:d_\alpha d_\beta K^{\alpha\beta}:\;+\;:d_\alpha N^{mn}G^\alpha_{\;\;mn}:\;+\;:d_\alpha J F^\alpha:\;+\;:d_\alpha\partial\theta^\beta F^\alpha_{\;\;\beta}:
\nonumber\\
&&+\;\;:N^{mn}N^{pq}G_{mnpq}:\;+\;:N^{mn}JP_{mn}:\;+\;:N^{mn}\partial\theta^\alpha H_{mn\alpha}:
\nonumber\\
&&+\;\;:JJH:\;+\;:J\partial\theta^\alpha H_\alpha :\;+\;:\partial\theta^\alpha\partial\theta^\beta H_{\alpha\beta}:
\label{general_U}
\ee
The terms in the first line involve derivatives of fields to produce objects of conformal weight 2. The terms in the last 4 lines involve products of fields with conformal weights 1 to produce objects of conformal weight 2. Note that the superfields contain the expansion in $\theta^\alpha$. Hence, there are no explicit $\theta^\alpha$ dependent terms in the above expression.

\vspace*{.06in}To set up the equation of motion \eqref{massive_for}, we now need to compute $QU$. Before stating the result, we note that the superfields appearing in \eqref{general_U} must be expressible in terms of the basic superfields $B_{mnp}, G_{mn}$ and $\Psi_{m\alpha}$. Moreover, we shall argue below that the superfields whose theta independent components can't contain the physical fields $b_{mnp}, g_{mn}$ and $\psi_{m\alpha}$ must be zero. These superfields are $C_\alpha, C_m, E^\alpha, C, C_{mn}, F_m, F^\alpha, P_{mn}, H $ and $H_\alpha$. Keeping this in mind, the action of the BRST operator $ Q$ on the 10 non zero terms of \eqref{general_U} can be computed to be\footnote{These computations were also checked using the Mathematica package OPEDefs \cite{opedefs2}. }

\begin{enumerate}
%\vspace*{.2in}\ul{$\partial^2\theta^\alpha C_\alpha$:}

\item 
\ul{$\Pi^m\Pi^nF_{mn}$}
\be 
Q\left(:\Pi^m\Pi^nF_{mn}:\right)
&=&\f{\alpha'}{2}\biggl[:\Pi^m \Pi^n\lambda^\alpha D_\alpha F_{mn}:\;+\;:\Pi^m (\gamma^n_{\alpha\beta})\p\theta^\beta\lambda^\alpha \Bigl(F_{mn}+F_{nm}\Bigl):\biggl]
\non
\ee

\item\ul{$\Pi^md_\alpha F_{m}^{\;\;\alpha}$}
\be 
Q\left(:\Pi^md_\beta F_{m}^{\;\;\beta}:\right)
&=&-\f{\alpha'}{2}\Bigl[:\Pi^m d_\beta\lambda^\alpha D_\alpha F_{m}^{\;\;\beta}:\;+\;: d_\beta (\gamma^m_{\alpha\sigma})\p\theta^\sigma\lambda^\alpha F_{m}^{\;\;\beta}:\non\\
&&+\;: \Pi^m (\gamma^n_{\alpha\beta})\Pi_n\lambda^\alpha F_{m}^{\;\;\beta}:\Bigl]
-\;\f{1}{2}\left(\f{\alpha'}{2}\right)^2\p^2\lambda^\alpha\gamma^m_{\alpha\sigma}F_{m}^{\;\;\sigma}\non\\
&&+\ \f{(\alpha')^2}{2}: \Pi^m(\gamma^n_{\alpha\beta})\p\lambda^\alpha
\p_nF_{m}^{\;\;\beta}:
\non
\ee

\item
\ul{$\Pi^mN^{pq}F_{mpq}$}
\be 
Q\left(:\Pi^mN^{pq}F_{mpq}:\right)
&=&\f{\alpha'}{2}\Bigl[:\Pi^mN^{pq}\lambda^\alpha D_\alpha F_{mpq}:\;+\;: \p\theta^\sigma N^{pq}(\gamma^m_{\alpha\sigma})\lambda^\alpha F_{mpq}:\Bigl]
\non\\
&&-\f{\alpha'}{4}\;: \Pi^m d_\alpha(\gamma^{pq})^\alpha_{\;\;\beta}\lambda^\beta F_{mpq}:-\f{1}{2}\left(\f{\alpha'}{2}\right)^2: \Pi^m\p\lambda^\beta
(\gamma^{pq})^\alpha_{\;\;\beta}D_\alpha F_{mpq}:
\non\\
&&-\;\f{1}{2}\left(\f{\alpha'}{2}\right)^2\Bigl[\p^2\theta^\sigma\lambda^\beta\gamma^m_{\alpha\sigma}(\gamma^{pq})^\alpha_{\;\;\beta}F_{mpq}+\p\theta^\sigma\p\lambda^\beta\gamma^m_{\alpha\sigma}(\gamma^{pq})^\alpha_{\;\;\beta}F_{mpq}\Bigl]
\non
\ee

\item 
\ul{$\Pi^m\p\theta^\beta G_{m\beta} $}
\be 
&&\hspace{-0.5in}Q\left(:\Pi^m\p\theta^\beta G_{m\beta}:\right)\non\\
&=&-\f{\alpha'}{2}:\Pi^m \p\theta^\beta\lambda^\alpha D_\alpha G_{m\beta}:\;+\;\f{\alpha'}{2}: \p\theta^\sigma\p\theta^\beta\lambda^\alpha \gamma^m_{\alpha\sigma}G_{m\beta}:\;+\;\f{\alpha'}{2}\;:\Pi^m\p\lambda^\beta G_{m\beta}:\non
\ee

\item
\ul{$ d_\alpha d_{\beta}K^{\alpha\beta}$}
\be 
Q\left(:d_\alpha d_\beta K^{\alpha\beta}:\right)
  &=&\f{\alpha'}{2}:d_\sigma d_\beta\lambda^\alpha D_\alpha K^{\sigma\beta}: {-\f{\alpha'}{2}:\Pi_m d_\beta(x)\lambda^\alpha\gamma^m_{\alpha\sigma} }
  \big[K^{\sigma\beta}(z) - K^{\beta\sigma}\big]:\non\\
  &&+\f{\alpha'^2}{2}:d_\beta \p\lambda^\alpha \gamma^m_{\alpha\sigma}\p_m\big[K^{\sigma\beta}- K^{\beta\sigma}\big]:
  +\;\left(\f{\alpha'}{2}\right)^2  \p\theta^\delta\p\lambda^\alpha\gamma_{m\beta\delta}\gamma^m_{\alpha\sigma}K^{\sigma\beta}\non\\
 && +\left(\f{\alpha'}{2}\right)^2:\gamma_{n\sigma\rho}\p^2\theta^\rho(x) \lambda^\alpha(z)\gamma^n_{\alpha\beta} K^{\sigma\beta}\non
\ee

\item
\ul{$d_\beta N^{mn}G^\beta _{mn}$}
\be 
&&\hspace{-0.5in}Q\left(:d_\beta N^{mn}G^\beta _{mn}:\right)\non\\
&=&\f{\alpha'}{2}\biggl[-:d_\beta N^{mn}{\lambda^\alpha}{D}_{\alpha}G^{\beta}_{mn}:-:\Pi^pN^{mn}\lambda^\alpha\gamma_{p\alpha\beta} G^\beta_{mn}:
+\alpha':N^{mn}\p\lambda^\alpha\gamma_{p\alpha\beta}\p^pG^\beta_{mn}:\non\\
&&+\f{\alpha'}{4}(\gamma_p\gamma^{mn})_{\beta\sigma}\Big(:\p \Pi^p\lambda^\sigma G^\beta_{mn} +:\Pi^p\p\lambda^\sigma G^\beta_{mn}:
-\f{\alpha'}{2} :\p^2\lambda^\sigma \p^pG^\beta_{mn}:\Big)\non\\
&&+\f{(\gamma^{mn})^\alpha_{\;\;\sigma}}{2}\Big(:d_\beta d_\alpha\lambda^\sigma G^\beta_{mn} :(z)
+\f{\alpha'}{2}:d_\beta \p\lambda^\sigma D_\alpha G^\beta_{mn} :\Big)\bigg]
\ee

\item
\ul{$d_\beta\p\theta^\delta F^{\beta}_{\;\;\delta}$}
\be 
Q\left(:d_\beta \p\theta^\delta F^{\beta}_{\;\delta}:\right)
&=&\f{\alpha'}{2}\left[:d_\beta \p\theta^\delta\lambda^\alpha D_\alpha F^{\beta}_{\;\;\delta}:\;-\;: d_\beta\p\lambda^\alpha F^{\beta}_{\;\;\alpha}:\right]{-\f{\alpha'}{2}: \Pi_m\p\theta^\delta \lambda^\alpha\gamma^m_{\alpha\beta}F^{\beta}_{\;\;\delta}:}\non\\
&&+\f{(\alpha')^2}{2}: \p\theta^\delta \p\lambda^\alpha\gamma^m_{\alpha\beta}\p_mF^{\beta}_{\;\;\delta}:
\non
\ee

\item
\ul{$N^{mn}N^{pq}G_{mnpq}$}
\be 
&&\hspace{-0.5in}Q\left(:N^{mn}N^{pq}G_{mnpq}:\right)\non\\
&=&\left(\f{\alpha'}{4}\right)^2\biggl[\f{8}{\alpha'}:N^{mn}N^{pq}\lambda^\alpha D_\alpha G_{mnpq}:-\f{4}{\alpha'}:d_\alpha N^{pq}\lambda^\beta (\gamma^{mn})^{\alpha}_{\;\;\beta} G_{mnpq}:
\non\\
&&-2:N^{pq}\p\lambda^\beta(\gamma^{mn})^{\alpha}_{\;\;\beta} D_\alpha G_{mnpq}:+(\gamma^{mn}\gamma^{pq})^\alpha_{\;\;\beta}\left(:\p d_\alpha\lambda^\beta G_{mnpq}:
+:d_\alpha \p\lambda^\beta G_{mnpq}:\right)\non\\
&&-\f{4}{\alpha'}:d_\alpha N^{mn}\lambda^\beta (\gamma^{pq})^{\alpha}_{\;\;\beta}G_{mnpq}:-2: N^{mn}\p\lambda^\beta (\gamma^{pq})^{\alpha}_{\;\;\beta}
D_\alpha G_{mnpq}:\non\\
&&+\f{\alpha'}{4}\Big( \p^2\lambda^\beta D_\alpha (\gamma^{mn}\gamma^{pq})^\alpha_{\;\;\beta}G_{mnpq}\Big)\biggl]
\ee

\item
\ul{$N^{mn}\p \theta^\beta H_{mn\beta}$}
\be
&&\hspace{-0.5in}Q\left(:\p\theta^\beta N^{mn}  H_{mn\beta}:\right)\non\\
&=&\f{\alpha'}{2}\biggl[-:\p\theta^{\beta}N^{mn}{\lambda^\alpha}{D}_{\alpha}H_{mn\beta}:+ :N^{mn}\p\lambda^\beta H_{mn\beta}:
-\f{\alpha'}{8}:\p^2\lambda^\alpha(\gamma^{mn})^{\beta}_{\;\;\alpha}H_{mn\beta}:\non\\
&&-\f{1}{2}: d_\alpha \p\theta^\beta \lambda^\sigma (\gamma^{mn})^\alpha_{\;\;\sigma}H_{mn\beta}:
+\f{\alpha'}{4}:\p\theta^\beta \p\lambda^\sigma (\gamma^{mn})^\alpha_{\;\;\sigma} D_\alpha H_{mn\beta}:\biggl](z)
\ee

\item
\ul{$\p\theta^\beta \p\theta^\delta H_{\beta\delta}$}
\be 
Q\left(:\p\theta^\beta \p\theta^\delta H_{\beta\delta}:\right)
&=&\f{\alpha'}{2}\Big[:\p\theta^\beta\p\theta^\delta \lambda^\alpha D_\alpha H_{\beta\delta}:
-:\p\theta^\beta\p\lambda^\delta \big(H_{\beta\delta}-H_{\delta\beta}\big):\Big]\non
\ee

\end{enumerate}

The BRST equation of motion also involves the world-sheet derivative of the unintegrated vertex operator, namely, $\p_{\mathbb{R}} V$. Making use of the equation \eqref{vertex_ansatz} and the operator identity \eqref{factor_2}, we obtain
\be
\partial_\mathbb{R} V&=&: \partial\theta^\beta\partial\lambda^\alpha B_{\alpha\beta}:\ +\ : \Pi^m\partial\lambda^\alpha H_{m\alpha}:\ +\;\;:\partial^2\theta^\alpha \lambda^\beta\left( B_{\beta\alpha}+\alpha'\gamma^m_{\sigma\alpha}\p_mC^\sigma_{\;\;\beta} \right):\non\\
&&+\ :\partial\theta^\beta \partial\theta^\delta \lambda^\alpha D_\delta B_{\alpha\beta}:
\ +\ :  \Pi^m\partial\theta^\beta\lambda^\alpha\big(2\partial_m B_{\alpha\beta}+D_\beta H_{m\alpha}\big):\ +\ :\partial d_\beta \lambda^\alpha C^\beta_{\;\;\alpha}:
\non\\
&&+\ :d_\beta\partial\lambda^\alpha  C^\beta_{\;\;\alpha}:\ +\ :d_\beta\partial\theta^\sigma\lambda^\alpha D_\sigma C^\beta_{\;\;\alpha}:\ +\ :2\Pi^md_\beta\lambda^\alpha \partial_m C^\beta_{\;\;\alpha}:
\ +\ :\partial\Pi^m \lambda^\alpha H_{m\alpha}:\non\\
&&+\ :2\Pi^m \Pi^n \lambda^\alpha \p_nH_{m\alpha}:\ 
+\ :\partial N^{mn} \lambda^\alpha F_{\alpha mn}:\ 
+\ :N^{mn}\partial\lambda^\alpha F_{\alpha mn}:                                                           \non\\
&&+\ :\partial\theta^\beta N^{mn} \lambda^\alpha D_\beta F_{\alpha mn}:\ +\ :2\Pi^pN^{mn} \lambda^\alpha \partial_p \;F_{\alpha mn}:
\label{partialV}
\ee
where, we have used
\be
:2d_\beta \Pi^m\lambda^\alpha\partial_m C^\beta_{\;\;\alpha}:&=&:2\Pi^md_\beta \lambda^\alpha \partial_m  C^\beta_{\;\;\alpha}:+\alpha':\gamma^m_{\beta\sigma}\p^2\theta^\sigma\lambda^\alpha\p_m C^\beta_{\;\;\alpha}:
\label{example_ref}
\ee

We now need to equate $QU$ and $\p_{\mathbb{R}} V$. A convenient way to do this is to compare the same basis elements in both sides. For the conformal weight 2 and ghost number 1 pure spinor objects (which appear in $QU$ and $\p_{\mathbb{R}} V$), naively, we have following 26 basis elements
\be
&&\Pi^m \Pi^n \lambda^\alpha \;,\;  \Pi^m d_\alpha  \lambda^\beta \;,\;  \Pi^m \p\theta^\beta  \lambda^\gamma \; ,\; \Pi^m J  \lambda^\alpha  \; ,\; \Pi^m N^{np}  \lambda^\alpha \;, \; \p\Pi^m  \lambda^\alpha \; , \; \Pi^{m}\p  \lambda^\alpha
\non\\ [.2cm]
&&\hspace*{.55in}d_\alpha d_\beta  \lambda^\gamma \; ,\; d_\alpha \p\theta^\beta  \lambda^\gamma \; ,\; d_\alpha J  \lambda^\alpha  \; , \; d_\alpha N^{mn}   \lambda^\alpha \;,\; \p d_\alpha  \lambda^\beta  \; , \; d_\alpha \p  \lambda^\beta
\non \\  [.2cm]
&&\hspace*{.65in}\p \theta^\alpha \p \theta^\beta \lambda^\gamma \; \;,\; 
 \p \theta^\alpha J \lambda^\beta \;,\; \p \theta^\alpha N^{mn} \lambda^\alpha \;, \; \p^2 \theta^\alpha \lambda^\beta \; , \; \p \theta^\alpha \p  \lambda^\beta 
 \non\\  [.2cm]
&&\hspace*{.9in}N^{mn} N^{pq}  \lambda^\alpha \;,\;  N^{mn} J  \lambda^\alpha \; , \; \p N^{mn}  \lambda^\alpha  \; ,\; N^{mn} \p  \lambda^\alpha
 \non\\  [.2cm]
 &&\hspace*{1.6in}JJ  \lambda^\alpha \; ,\;  \p J  \lambda^\alpha \; , \; J\p  \lambda^\alpha
 \non \\ [.2cm] 
&&\hspace*{2.1in}
 \p^2 \lambda^\alpha 
\ee

As mentioned earlier, all of these basis elements are not independent. There are non trivial relations among some of these bases. We turn to these constraint relations between the basis elements in the next subsection. 

\subsection{Constraint Identities }
\label{Constraint_iden3}
As mentioned in section \ref{sec:review}, due to pure spinor constraint, the Lorentz current $N^{mn}$ and the ghost current $J$ satisfy the identity \cite{Berkovits1}
\be
:N^{mn}\lambda^\alpha :(z)(\gamma_m)_{\alpha\beta}-\f{1}{2}:J\lambda^\alpha :(z)(\gamma^n)_{\alpha\beta}-\alpha'\gamma^n_{\alpha\beta}\p\lambda^\alpha(z)\ =\ 0
\label{identity123}
\ee
This constraint is relevant if one is interested in the quantities involving conformal weight 1 and ghost number 1. However, in the expressions for $QU$ and $\p_{\mathbb{R}} V$, we encounter quantities with conformal weight 2 and ghost number 1. For this case, there are several identities which can be obtained from the above identity \eqref{identity} by taking the OPE of this with the objects of conformal weight 1 and demanding the normal order terms in the OPE to vanish (the pole terms of the OPE vanish automatically as expected). Since the derivative and the normal ordering commute, the world-sheet derivative of \eqref{identity123} also gives a constraint. We list these constraint identities below. 
\be
(I_1)_\beta^n&\equiv&:N^{mn}J\lambda^\alpha :(\gamma_m)_{\alpha\beta}-\f{1}{2}:JJ\lambda^\alpha :(\gamma^n)_{\alpha\beta}-\alpha':J\p\lambda^\alpha:\gamma^n_{\alpha\beta}\ =\ 0
\label{identity1}
\\[.3cm]
(I_2)_\beta^{mnq}&\equiv&:N^{mn}N^{pq}\lambda^\alpha :(\gamma_p)_{\alpha\beta}-\f{1}{2}:N^{mn}J\lambda^\alpha :(\gamma^q)_{\alpha\beta}-\alpha':N^{mn}\p\lambda^\alpha:\gamma^q_{\alpha\beta}\ =\ 0
\label{identity2}
\\[.3cm]
(I_3)_{\sigma\beta}^{ n}&\equiv&:d_\sigma N^{mn}\lambda^\alpha :(\gamma_m)_{\alpha\beta}-\f{1}{2}:d_\sigma J\lambda^\alpha :(\gamma^n)_{\alpha\beta}-\alpha':d_\sigma\p\lambda^\alpha:\gamma^n_{\alpha\beta}\ =\ 0
\label{identity3}
\\[.3cm]
(I_4)_\beta^{pn}&\equiv&:\Pi^pN^{mn}\lambda^\alpha :(\gamma_m)_{\alpha\beta}-\f{1}{2}:\Pi^pJ\lambda^\alpha :(\gamma^n)_{\alpha\beta}-\alpha':\Pi^p\p\lambda^\alpha:\gamma^n_{\alpha\beta}\ =\ 0
\label{identity4}
\\[.3cm]
(I_5)_\beta^{\sigma n}&\equiv&:\p\theta^\sigma N^{mn}\lambda^\alpha :(\gamma_m)_{\alpha\beta}-\f{1}{2}:\p\theta^\sigma J\lambda^\alpha :(\gamma^n)_{\alpha\beta}-\alpha':\p\theta^\sigma\p\lambda^\alpha:\gamma^n_{\alpha\beta}\ =\ 0
\label{identity5}
\ee
The above 5 identities follow from taking the OPE of \eqref{identity123} with the object of conformal weight one, namely $J, N^{mn}, d_\sigma, \Pi^p$ and $\p\theta^\sigma$ respectively.
The identity which can be obtained by taking the derivative of \eqref{identity123} is given by
\be
(I_6)_\beta^{ n}&\equiv&:\p N^{mn}\lambda^\alpha :(\gamma_m)_{\alpha\beta}+ :N^{mn}\p\lambda^\alpha :(\gamma_m)_{\alpha\beta}-\f{1}{2}:\p J\lambda^\alpha :(\gamma^n)_{\alpha\beta}-\f{1}{2}:J\p\lambda^\alpha :(\gamma^n)_{\alpha\beta}\non\\
&&\hspace*{.04in}-\ \alpha'\gamma^n_{\alpha\beta}\p^2\lambda^\alpha\ =\ 0\label{identity6}
\ee
Apart from these, there are two more constraint identities which follow from the OPEs given in section \ref{sec:review}. The OPE of $d_\alpha$ with $d_\beta$ implies
\be
:d_\alpha d_\beta:\ +\ :d_\beta d_\alpha:\ +\ \f{\alpha'}{2}\p\Pi^t (\gamma_{t})_{\alpha\beta}\ =\ 0\label{dalphadbeta}
\ee
Similarly, the OPE of $N^{mn}$ with $N^{pq}$ implies
\be
:N^{mn}N^{pq}:\ -\ :N^{pq}N^{mn}:\ =\ -\f{\alpha'}{2}\Bigl[\eta^{np}\p N^{mq}-\eta^{nq}\p N^{mp}-\eta^{mp}\p N^{nq}+\eta^{mq}\p N^{np}\Bigl]\label{NmnNpqiden}
\ee
One way to think about these two identities is to note that we are working with a given ordering of the pure spinor variables inside the normal ordering. However, for $:d_\alpha d_\beta:$ and $:N^{mn} N^{pq}:$, there is no preferred ordering. The above two identities \eqref{dalphadbeta} and \eqref{NmnNpqiden} are a reflection of this fact\footnote{Note that there are OPE between $\Pi^m$ and $\Pi^n$ as well as between $J$ and $J$. However, no pure spinor fields appear in these OPE and hence they do not lead to any non trivial constraint between basis elements.}. 

\vspace*{.06in}For later purpose, we multiply \eqref{dalphadbeta} with 5-form $\gamma_{mnpqr}^{\alpha\beta}$ to obtain
\be
\gamma_{mnpqr}^{\alpha\beta}\Bigl(:d_\alpha d_\beta: +:d_\beta d_\alpha: +\f{\alpha'}{2}\p\Pi^t (\gamma_{t})_{\alpha\beta}\Bigl)=0
\qquad\implies \quad  \gamma_{mnpqr}^{\alpha\beta}:d_\alpha d_\beta:\ =\ 0\label{ddcons345}
\ee 
where, we have used the fact that the trace of product of 5-form and 1-form is zero and the 5-form is symmetric in its spinor indices.

\vspace*{.06in}For solving the equations of motion, we shall need to take into account all of these constraint relations between the pure spinor variables.

\subsection{Setting up the Equations}
\label{eqsetup}
We shall now equate $QU$ and $\p_{\mathbb{R}} V$ and solve the resulting equations of motion. As mentioned earlier, a convenient way to do this is to equate the terms with the same basis elements taking into account the constraint identities given above. 

\vspace*{.06in}To take into account the constraint identities, we have two options - eliminate some basis in terms of others or introduce Lagrange multipliers. 
We shall make use of both of these options. We shall use the elimination method for taking care of \eqref{dalphadbeta} and \eqref{NmnNpqiden} constraints. More specifically, we shall eliminate the basis involving $\p \Pi^m$ in favour of the basis involving $d_\alpha d_\beta$ and similarly we shall eliminate the anti-symmetric part of the basis involving $N^{mn}N^{pq}$ (in simultaneous $m\leftrightarrow p$ and $n\leftrightarrow q$ exchange) in the favor of basis involving $\p N^{mn}$. On the other hand, we shall introduce Lagrange multipliers for the six constraints \eqref{identity1}-\eqref{identity6} which follow from the pure spinor constraint and involve the pure spinor ghost. This means that we add a very specific zero to $QU=\p_{\mathbb{R}} V$ equation so that we have 
\be 
QU=\p_{\mathbb{R}} V + \sum_{a=1}^{6} I_{a}K_a\label{eqnBRST}
\ee
The $I_aK_a$ involve contraction of the six identities \eqref{identity1}-\eqref{identity6} with appropriate Lagrange multiplier superfields. We denote these arbitrary superfields by $K_i \;\;(i=1,\cdots 6)$. Thus, 
\be
 \sum_{a=1}^{6} I_{a}K_a&\equiv& (I_1)_\beta^n (K_1)^\beta_n\ +\
(I_2)_\beta^{mnq}(K_2)^\beta_{mnq}\ +\
(I_3)_{\sigma\beta}^{ n}(K_3)^{\sigma\beta}_{ n}\ +\ (I_4)_\beta^{pn}(K_4)^\beta_{pn}\non\\[.1cm]
&&
+\
(I_5)_\beta^{\sigma n}(K_5)^\beta_{\sigma n}\ +\ 
(I_6)_\beta^{ n}(K_6)^\beta_{ n}\label{form_of_L}
\ee
The Lagrange multiplier superfields $K_i$ will also be determined in terms of the basic superfields $B_{mnp}, G_{mn}$ and $\Psi_{m\alpha}$ as we shall see.

\vspace*{.09in} We can now write down the equations of motion. Using equations \eqref{form_of_L}, \eqref{identity1}-\eqref{identity6} and the expressions of $QU$ and $\p_{\mathbb{R}} V$, we obtain the following equations after comparing the same basis elements in both sides of \eqref{eqnBRST}
\begin{enumerate}

%7%
\item $\ul{\Pi^m\Pi^n \lambda^\alpha}$
\be
\f{\alpha'}{2}\biggl[D_\alpha F_{mn}-\gamma_{n\alpha\beta}F_m^{\;\;\beta}\biggl]= 2\p_n H_{m\alpha}
\non
\ee

%8%
\item $\ul{\Pi^m\p\theta^\beta \lambda^\alpha}$
\be
\f{\alpha'}{2}\biggl[\gamma^n_{\alpha\beta}(F_{mn}+F_{nm})-D_\alpha G_{m\beta} {-\gamma^m_{\alpha\delta}F^{\delta}_{\;\;\beta}}\biggl]=2\partial_m B_{\alpha\beta}+D_\beta H_{m\alpha}
\non
\ee

%12%
\item $\ul{d_\alpha \p\theta^\beta \lambda^\sigma}$
\be
\f{\alpha'}{2}\biggl[-\gamma^m_{\sigma\beta}F_m^{\;\;\alpha}+D_\sigma F^\alpha_{\;\;\beta}
-\f{1}{2}(\gamma^{mn})^{\alpha}_{\;\;\sigma}H_{mn\beta}\biggl]=D_{\beta}C^\alpha_{\;\;\sigma}
\non
\ee

%13%
\item $\ul{\Pi^md_\beta \lambda^\alpha}$
\be
\f{\alpha'}{2}\biggl[-D_\alpha F_m^{\;\;\beta}-\f{1}{2}(\gamma^{pq})^\beta_{\;\;\alpha}F_{mpq} 
-\gamma^m_{\alpha\sigma} \Big(K^{\sigma\beta} - K^{\beta\sigma}\Big) \biggl]={2
\p_m C^\beta_{\;\;\alpha}}
\non
\ee

%15%
\item $\ul{\p\theta^\alpha\p\theta^\beta \lambda^\sigma}$
\be
\f{\alpha'}{2}\biggl[\gamma^m_{\sigma[\alpha}G_{m\beta]}+ D_\sigma H_{\alpha\beta}\biggl]&=& D_{[\beta} B_{|\sigma|\alpha]}
\non
\ee

%1%
\item $\ul{\p\Pi_m\lambda^\alpha}$
\be
\f{(\alpha')^2}{8}(\gamma_m\gamma^{pq})_{\beta\alpha} G^\beta_{pq}
=H_{m\alpha}
\non
\ee

\item $\ul{d_\alpha d_\beta \lambda^\sigma}$
\be
\f{\alpha'}{2}\biggl[D_\sigma K^{\alpha\beta}+\f{1}{2}(\gamma^{mn})^\beta_{\;\;\sigma}G_{mn}^\alpha
\biggl]=0
\non
\ee

%3%
\item $\ul{\p^2\theta^\beta \lambda^\alpha}$
\be
\f{\alpha'}{2}\biggl[-\f{\alpha'}{4}\gamma^m_{\beta\sigma}(\gamma^{pq})^\sigma_{\;\;\alpha} F_{mpq}
+\f{\alpha'}{2}\gamma^m_{\delta\beta}\gamma_{m\alpha\sigma}K^{{\delta\sigma}}\biggl]= B_{\alpha\beta}+\alpha'\gamma^m_{\sigma\beta}\p_mC^\sigma_{\;\;\alpha}
\non
\ee

%10%
\item $\ul {\Pi^mN^{pq} \lambda^\alpha}$
\be
\f{\alpha'}{2}\biggl[D_\alpha F_{mpq}-\gamma_{m\alpha\beta}G^\beta_{\;\;pq}\biggl]= 2\p_m F_{\alpha pq}+(\gamma_{[p})_{\alpha\beta}(K_4)^\beta_{\;\;|m|q]}
\non
\ee

%9%
\item $\ul{\Pi^mJ \lambda^\alpha}$
\be
0=-\f{1}{2}\gamma^{q}_{\;\;\alpha\beta}(K_4)^\beta_{\;\;mq}
\non
\ee

%11%
\item $\ul{\Pi^m\p \lambda^\alpha}$
\be
&&\f{\alpha'}{2}\biggl[\alpha'\gamma^n_{\alpha\beta}\p_n F_m^{\;\;\beta}-\f{\alpha'}{4}(\gamma^{pq})^\beta_{\;\;\alpha}D_\beta F_{mpq}
+G_{m\alpha} +\f{\alpha'}{4}(\gamma_m\gamma^{pq})_{\beta\alpha} G^\beta_{pq}\biggl]
\non\\[.3cm]
&&\hspace*{2.5in}= H_{m\alpha}-\alpha'\gamma^{q}_{\alpha\beta}(K_4)^\beta_{\;\;mq}
\non
\ee

%17%
\item $\ul{\p\theta^\alpha N^{mn} \lambda^\beta}$
\be
\f{\alpha'}{2}\biggl[\gamma^p_{\alpha\beta}F_{pmn}-D_\beta H_{mn\alpha}\biggl]&=& D_\alpha F_{\beta mn}+(\gamma_{[m})_{\beta\sigma}(K_5)^{\sigma}_{\;\;\alpha n]}
\non
\ee

%16%
\item $\ul{\p\theta^\alpha J\lambda^\beta}$
\be
0\ =\ -\f{1}{2}\gamma^n_{\beta\sigma}(K_5)^{\sigma}_{\;\;\alpha n}
\non
\ee

%18%
\item $\ul{\p\theta^\alpha\p \lambda^\beta}$
\be
&&\hspace*{-.46in}\f{\alpha'}{2}\biggl[-\f{\alpha'}{4}\gamma^m_{\alpha\sigma}(\gamma^{pq})^\sigma_{\;\;\beta} F_{mpq}
+\f{\alpha'}{2}\gamma^m_{\delta\alpha}\gamma_{m\beta\sigma}K^{\sigma\delta}
+\alpha'\gamma^m_{\beta\sigma}\p_mF^\sigma_{\;\;\alpha}+\f{\alpha'}{4}(\gamma^{mn})^\sigma_{\;\;\beta}D_{\sigma}H_{mn\alpha}- 2H_{\alpha\beta}\biggl]\non\\[.5cm]
&&\hspace*{3.3in}=B_{\beta\alpha}-\alpha' \gamma^n_{\beta\sigma}(K_5)^{\sigma}_{\;\;\alpha n}
\non
\ee

%6%
\item \ul{$\p^2 \lambda^\alpha$}
\be
&&\hspace*{-.3in}\f{\alpha'}{2}\biggl[-\f{\alpha'}{4}\gamma^m_{\alpha\beta}F_m^{\;\;\beta}-\f{(\alpha')^2}{8}(\gamma_m\gamma^{pq})_{\beta\alpha} \p^mG^\beta_{pq}
+\f{\alpha'^2}{32}(\gamma^{mn}\gamma^{pq})^\beta_{\;\;\alpha}D_\beta G_{mnpq}
-\f{\alpha'}{8}(\gamma^{mn})^\beta_{\;\;\alpha}H_{mn\beta} \biggl]\non\\[.3cm]
&&\hspace*{4.5in}=  -\alpha' \gamma^n_{\alpha\beta}(K_6)_n^\beta
\non
\ee

%4%
\item $\ul{\p J \lambda^\alpha}$
\be
0= -\f{1}{2} \gamma^n_{\alpha\beta}(K_6)_n^\beta
\non
\ee

%19%
\item $\ul{J\p\lambda^\alpha}$
\be
0=-\alpha'\gamma^{n}_{\;\alpha\beta}(K_1)_n^\beta-\f{1}{2}\gamma^n_{\alpha\beta}(K_6)_n^\beta
\non
\ee

\item $\ul{JJ\lambda^\alpha}$
\be
0\ =\ -\f{1}{2}\gamma^{n}_{\;\alpha\beta}(K_1)_n^\beta
\non
\ee

%2%
\item $\ul{\p d_\alpha\lambda^\beta}$
\be
\f{(\alpha')^2}{16}(\gamma^{mn}\gamma^{pq})^\alpha_{\;\;\beta}G_{mnpq}= C^{\alpha}_{\;\;\beta}
\non
\ee

 \item $\ul{d_\alpha N^{mn} \lambda^\beta}$
 \be
\f{\alpha'}{2}\biggl[-D_\beta G^\alpha_{mn}-\f{1}{2}(\gamma^{pq})^{\alpha}_{\;\;\beta}\Big(G_{mnpq}+G_{pqmn}\Big)
\biggl]=(\gamma_{[m})_{\beta\sigma}(K_3)^{\sigma\alpha}_{\;\;n]}
\non
\ee

\item $\ul{d_\alpha J \lambda^\beta}$
\be
0
\ =\ -\f{1}{2}\gamma^n_{\;\;\beta\sigma}(K_3)^{\sigma\alpha}_{\;\;n}
\non
\ee

%14%
\item $\ul{d_\alpha \p \lambda^\beta}$
\be
&&\f{\alpha'}{2}\biggl[\alpha'\gamma^n_{\beta\sigma}(\p_n K^{\sigma\alpha}-\p_n K^{\alpha\sigma})+\f{\alpha'}{4}(\gamma^{mn})^\sigma_{\;\;\beta}D_\sigma G^\alpha_{mn}
-F^\alpha_{\;\;\beta}+\f{\alpha'}{8}(\gamma^{mn}\gamma^{pq})^\alpha_{\;\;\beta}G_{mnpq}
\biggl]\non\\[.3cm]
&&\hspace*{3.3in}=
C^\alpha_{\;\;\beta}-\alpha'\gamma^n_{\;\;\beta\sigma}(K_3)^{\sigma\alpha}_{\;\;n}
\non
\ee

%20%
\item $\ul{N^{mn}\p\lambda^\alpha}$
\be
&&\hspace*{-.6in}\f{\alpha'}{2}\biggl[\alpha'\gamma_{p\alpha\beta}\p^pG^\beta_{\;\;mn}-\f{\alpha'}{4}(\gamma^{pq})^{\beta}_{\;\;\alpha} D_\beta 
\Big(G_{mnpq}+G_{pqmn}\Big)+H_{mn\alpha}\biggl]
\non\\[.3cm]
&=& F_{\alpha mn}-\alpha' \gamma^{q}_{\alpha\beta}(K_2)^{\beta}_{\;\;mnq}+(\gamma_{[m})_{\alpha\beta}(K_6)_{n]}^\beta
\non
\ee

\item $\ul{JN^{mn}\lambda^\alpha}$
\be
0\ =\ (\gamma_{[m})_{\alpha\beta}(K_1)_{n]}^\beta-\f{1}{2}\gamma^{q}_{\alpha\beta}(K_2)^{\beta}_{\;\;mnq}
\non
\ee

%5%
\item $\ul{\p N^{mn} \lambda^\alpha}$
\be
0= F_{\alpha mn}+(\gamma_{[m})_{\alpha\beta}(K_6)_{n]}^\beta
\non
\ee

\item $\ul{N^{mn}N^{pq}\lambda^\alpha}$
\be
\f{\alpha'}{2}\biggl[D_\alpha G_{mnpq}\biggl]=(\gamma_{[p})_{\alpha\beta}(K_2)^{\beta}_{\;\;|mn|q]}
\non
\ee

\end{enumerate}
We have not yet taken into account the constraints imposed by \eqref{dalphadbeta} and \eqref{NmnNpqiden} on the basis elements. We do this now and first consider \eqref{dalphadbeta} which will relate 6th and the 7th equations of the above 26 equations. Eliminating $\p \Pi^m$ in 6th equation in favor of $d_\alpha d_\beta$ using \eqref{dalphadbeta} and combining it with the 7th equation gives following equation for the coefficient of $d_\alpha d_\beta\lambda^\sigma$ 
\be
\f{\alpha'}{2}\biggl[D_\sigma K^{\alpha\beta}-\f{1}{2}(\gamma^{mn})^\alpha_{\;\;\sigma}G_{mn}^\beta-\f{36}{(\alpha')^2}\gamma^{\alpha\beta}_m\Psi^m_{\sigma}
\biggl]=0
\label{combined6th7th}
\ee
Next, we consider \eqref{NmnNpqiden} which relates the basis involving $\p N^{mn}$ with the anti symmetric part of the basis involving $N^{mn}N^{pq}$. This will relate 25th and the 26th equations. We first seperate the symmetric and the anti symmetric parts of $N^{mn}N^{pq}$ of 26th equation and then combine the anti symmetric part with 25th equation using \eqref{NmnNpqiden}. 

\vspace*{.06in}The anti symmetric part of $QU-\p_{\mathbb{R}} V-\sum_i I_i$ side of the 26th equation is given by
\be
&&\f{1}{2}:\Bigl(N^{mn}N^{pq}-N^{pq}N^{mn}\Bigl)\lambda^\alpha\biggl[\f{\alpha'}{2}D_\alpha G_{mnpq}-(\gamma_{p})_{\alpha\beta}(K_2)^{\beta}_{\;\;mnq}\biggl]:\non\\
&=&-\f{\alpha'}{2}\p N^{mn}\lambda^\alpha\biggl[\alpha'\eta^{pq}D_\alpha G_{mpqn}+(\gamma^{p})_{\alpha\beta}(K_2)^{\beta}_{\;\;pmn}-\eta^{pq}(\gamma_{m})_{\alpha\beta}(K_2)^{\beta}_{\;\;npq}\biggl]\non
\ee
where, we have used equation \eqref{NmnNpqiden} in going from the first to second line. 

\vspace*{.06in}Combining this with the 25th equation and demanding the coefficient of $\p N^{mn}\lambda^\alpha$ to vanish gives the following equation
\be
&&\f{\alpha'}{2}\biggl[-\alpha'\eta^{pq}D_\alpha G_{[m|pq|n]}-(\gamma^{p})_{\alpha\beta}(K_2)^{\beta}_{\;\;p[mn]}+\eta^{pq}(\gamma_{[m})_{\alpha\beta}(K_2)^{\beta}_{\;\;n]pq}\biggl]-F_{\alpha mn}-(\gamma_{[m})_{\alpha\beta}(K_6)_{n]}^\beta\non\\
&&\hspace*{5in}=0\label{asym26and5}
\ee
On the other hand, the symmetric part of $QU-\p_{\mathbb{R}} V-\sum_i I_i$ side of the 26th equation is given by 
\be
&&
\f{1}{2}:\Bigl(N^{mn}N^{pq}+N^{pq}N^{mn}\Bigl)\lambda^\alpha\biggl[\f{\alpha'}{2}D_\alpha G_{mnpq}-(\gamma_{p})_{\alpha\beta}(K_2)^{\beta}_{\;\;mnq}\biggl]:
\non\\
&=&
\f{1}{2}:N^{mn}N^{pq}\lambda^\alpha\biggl[\f{\alpha'}{2}\Bigl(D_\alpha G_{mnpq}+D_\alpha G_{pqmn}\Bigl)-(\gamma_{p})_{\alpha\beta}(K_2)^{\beta}_{\;\;mnq}-(\gamma_{m})_{\alpha\beta}(K_2)^{\beta}_{\;\;pqn}\biggl]:
\non
\ee
Demanding the coefficient of $N^{mn}N^{pq}\lambda^\alpha$ to vanish gives the following equation
\be
\f{\alpha'}{2}\Bigl(D_\alpha G_{mnpq}+D_\alpha G_{pqmn}\Bigl)-(\gamma_{[p})_{\alpha\beta}(K_2)^{\beta}_{\;\;|mn|q]}-(\gamma_{[m})_{\alpha\beta}(K_2)^{\beta}_{\;\;|pq|n]} =0
\label{sym26}
\ee
Our goal now is to find the superfields (and Lagrange multipliers) which satisfy the 26 equations listed earlier (except 5, 6, 25 and 26) and \eqref{combined6th7th}, \eqref{asym26and5} and \eqref{sym26}. If our superfields satisfy these equations, then they will  automatically satisfy the BRST equation of motion $QU=\p_{\mathbb{R}} V$.

\subsection{The ansatz for various superfields}

The equations of motion arising from $QU =\p_\mathbb{R} V$, in general, are very complicated due to the presence of gamma matrices and the super covariant derivatives. A direct approach based on comparing the different theta components of the superfields soon becomes messy and intractable. Due to this reason, we shall follow an alternative approach in which we directly propose an ansatz for the superfields and verify that they indeed satisfy the equations given in the previous section. \mvnote{These ansatz follow from the requirement of Lorentz invariance, equations of motion given in \eqref{D_Gmn1}-\eqref{cons_theta=0} and demanding that the superfields appearing in the integrated vertex should be expressible in terms of the 3 basic superfields $B_{mnp}, G_{mn} $ and $\Psi_{m\alpha}$. This allows us to work with the full covariant superfields instead of working with their theta components as required by the presence of super covariant derivatives. More details about how to arrive at these ansatz in given in appendix \ref{motive}.} 

\vspace*{.06in}Our proposed ansatz for expressing various superfields appearing in the integrated vertex in terms of the 3 basic superfields $B_{mnp}, G_{mn} $ and $\Psi_{m\alpha}$ and a set of unknown constant co-efficients are as follows 
\be
C_\alpha&=&C_m=E^\alpha=C=C_{mn}=F_m=F^\alpha=P_{mn}=H=H_\alpha=0\non\\[.4cm]
F_{mn}&=&f_1G_{mn}\qquad\qquad\qquad \qquad\;\;, \qquad G_{m\alpha}\ =\ g_1\Psi_{m\alpha} \non\\[.4cm]
 K^{\alpha \beta}&=& a\; \gamma_{mnp}^{\alpha \beta}B^{mnp}\qquad\qquad\qquad\;,\qquad H_{\alpha \beta}\ =\ h_1\gamma^{mnp}_{\alpha \beta}B_{mnp}  \non\\[.4cm]
F^{\alpha}_{\;\;\;\beta}&=&f_5 (\gamma^{mnpq})^{\alpha}_{\;\;\;\beta}k_mB_{npq}\qquad\quad\;,\qquad F_m^\alpha\ =\ f_2 k^r(\gamma_r)^{\alpha\beta}\Psi_{m\beta}
\non\\[.4cm]
 F_{mpq}&=&f_3G_{m[p}k_{q]}+f_4B_{mpq}\quad\quad\quad\;\;,\qquad G^{ \beta}_{pq}\ =\ g_2\gamma_{[p}^{ \beta\sigma}\Psi_{q]\sigma}+g_3 k^r\gamma_r^{ \beta\sigma}k_{[p}\Psi_{q]\sigma}
\non\\[.4cm]
 H_{mn\alpha}&=&h_2\; k_{[m}\Psi_{n]\alpha}+h_3k^q(\gamma_{q[m})_\alpha^{\;\;\sigma}\Psi_{n]\sigma}\non\\[.4cm]
G_{mnpq}&=&g_4k_{[m}B_{n]pq}+g_5k_{[p}B_{q]mn}+g_6k_{[m}G_{n][p}k_{q]}+g_7\;\eta_{[m[p}G_{{q]}n]}\label{Group_theory_U}
\ee
We also need similar ansatz for the Lagrange multipliers in terms of the basic superfields. We propose
\be
 (K_1)^\alpha_m&=&c_1 k^r(\gamma_r)^{\alpha\beta}\Psi_{m\beta}\non\\[.4cm]
(K_2)^\alpha_{mnq}&=&c_2k_{[m}\gamma_{n]}^{\alpha\beta}\Psi_{q\beta}+c_3k_{q}\gamma_{[m}^{\alpha\beta}\Psi_{n]\beta}
+c_4\gamma_{q}^{\alpha\beta}k_{[m}\Psi_{n]\beta}+c_5k^r\gamma_{rmn}^{\alpha\beta}\Psi_{q\beta}
+c_6k^r\gamma_{rq[m}^{\alpha\beta}\Psi_{n]\beta}\non\\[.2cm]
&&\hspace*{.01in}+\ c_7k^rk_q\gamma_{r}^{\alpha\beta}k_{[m}\Psi_{n]\beta}
+c_8k^r\gamma_{r}^{\alpha\beta}\eta_{q[m}\Psi_{n]\beta}\non\\[.4cm]
(K_3)^{\alpha\beta}_m&=&c_9 G_{mn}(\gamma^n)^{\alpha\beta}+c_{10} k_mB_{stu}(\gamma^{stu})^{\alpha\beta}+c_{11} k_sB_{tum}(\gamma^{stu})^{\alpha\beta}+c_{12} k_sB_{tuv}(\gamma^{\;\;\;stuv}_{m})^{\alpha\beta}\non\\[.4cm]
(K_4)^{\alpha}_{mn}&=&c_{13} (\gamma_n)^{\alpha\beta}\Psi_{m\beta}+c_{14} (\gamma_m)^{\alpha\beta}\Psi_{n\beta}+c_{15} k^rk_m(\gamma_{r})^{\alpha\beta}\Psi_{n\beta} +c_{16}k^rk_n(\gamma_{r})^{\alpha\beta}\Psi_{m\beta} \non\\[.4cm]
(K_5)^{\alpha}_{\;\;\beta m}&=&c_{17} k_{p}G_{qm}(\gamma^{pq})^{\alpha}_{\;\;\beta }+c_{18} B_{mpq}(\gamma^{pq})^{\alpha}_{\;\;\beta }+c_{19} B_{pqr}(\gamma^{\;\;\;pqr}_{m})^{\alpha}_{\;\;\beta }+c_{20}k_m k_pB_{qrs}(\gamma^{pqrs})^{\alpha}_{\;\;\beta }\non\\[.4cm]
(K_6)^\alpha_m&=&c_{21} k^r(\gamma_r)^{\alpha\beta}\Psi_{m\beta}\label{Lag_multi_ansatz}
\ee
Our job has now reduced to finding the unknown coefficients appearing in above ansatz. If we put these ansatz for the superfields in the equation of motion given above, we shall obtain a system of linear algebraic equations for the unknown coefficients which are much easier to solve. However, before doing this, we shall now see that there are some restriction on some of the coefficients which follow from the constraint identities given earlier and also directly from pure spinor condition. 

\vspace*{.06in}We start by noting that the superfield $G_{mnpq}$ appears in the expression of the integrated vertex operator as $N^{mn}N^{pq}G_{mnpq}$. We want to find the consequence of the identity \eqref{NmnNpqiden} on $G_{mnpq}$. For this, we consider the quantity $(N^{mn}N^{pq}-N^{pq}N^{mn})G_{mnpq}$. Using the identity \eqref{NmnNpqiden} and the ansatz for $G_{mnpq}$ given in \eqref{Group_theory_U}, we find that the right hand side of the identity \eqref{NmnNpqiden} vanishes identically after contraction with $G_{mnpq}$ and hence
\be
:(N^{mn}N^{pq}-N^{pq}N^{mn})G_{mnpq}:\ =0\quad\implies \qquad  :N^{mn}N^{pq} (G_{mnpq}-G_{pqmn}): \ = \ 0
\ee
This shows that $G_{mnpq}$ is symmetric under the exchange of simultaneous $m\leftrightarrow p$ and $n\leftrightarrow q$ indices. Now, the last two terms in the expression of $G_{mnpq}$ are already consistent with this property. However, this is not the case with the first two terms for which the tensor structures multiplying the coefficients $g_4$ and $g_5$ get exchanged. Thus, for $G_{mnpq}$ to be symmetric under the exchange of $m\leftrightarrow p$ and $n\leftrightarrow q$ indices, we must have
\be
g_4=g_5\label{g4=g5}
\ee

\vspace*{.06in}Next, we show that the term involving $g_7$ in the $G_{mnpq}$ superfield vanishes identically. For this, we first note that the term involving $g_7$ appears in the integrated vertex operator as
\be 
g_7 N^{mn}N^{pq}  \eta_{mp}G_{qn}=-g_7 N^{mn}N^{nq}G_{mq}\label{vanishg7}
\ee
Using the definition of $N^{mn}$, we obtain classically
\be 
N^{mn}N^{nq}G_{mq}&=&\f{1}{4}w_\alpha w_\sigma (\gamma^{mn})^{\alpha}_{\;\;\beta} (\gamma^{nq})^{\sigma}_{\;\;\rho} \lambda^\beta \lambda^\rho G_{mq}\non
%&=&\f{1}{4 \times 5! \times 16}w_\alpha w_\sigma (\gamma^{mn})^{\alpha}_{\;\;\beta} (\gamma^{nq})^{\sigma}_{\;\;\rho} (\gamma^{stuvw})^{\beta \rho}(\lambda\gamma_{stuvw} \lambda)G_{mq}\non\\&=&-\f{1}{4 \times 5! \times 16}w_\alpha w_\sigma (\gamma^{mn} \gamma^{stuvw}\gamma^{nq})^{\alpha \sigma}(\lambda\gamma_{stuvw} \lambda)G_{mq}\non\\&=&-\f{1}{4 \times 5! \times 16}w_\alpha w_\sigma\left[-\eta^{mq} \gamma^{stuvw}(\lambda\gamma_{stuvw} \lambda) -2 \gamma^{stuvwmq}(\lambda\gamma_{stuvw} \lambda)+ 40 \gamma^{stu} (\lambda\gamma_{stumq} \lambda)\right]^{\alpha \sigma} \non
\ee
The right hand side vanishes after using the fierz relation (which follows from the pure spinor condition)
\be
\lambda^\beta \lambda^\rho = \f{1}{32\times 5!}(\lambda\gamma_{stuvw}\lambda) \gamma_{stuvw}^{\beta\rho}\;\;\; ,
\ee
 the identities involving the product of gamma matrices and the symmetry and tracelessness properties of $G_{mq}$. We shall now show that this holds true even at the quantum level. The normal ordering piece which arises at quantum level is given by the right hand side of the identity \eqref{NmnNpqiden} contracted with $\eta^{np}$. So that the quantum version of the classical equation $N^{mn}N^{pq}\eta_{np}G_{mq}=0$ is given by
\be
:N^{mn}N^{pq}\eta_{np}G_{mq}:\ =\ c :\Bigl[\eta^{np}\p N^{mq}-\eta^{nq}\p N^{mp}-\eta^{mp}\p N^{nq}+\eta^{mq}\p N^{np}\Bigl]\eta_{np}G_{mq}:
\ee
where $c$ is an arbitrary coefficient which needs to be determined.  But, a little algebra shows that the right hand side is proportional to $
:\p N^{mq}G_{mq}:$ which is zero identically (since $N^{mq}$ is anti symmetric whereas $G_{mq}$ is symmetric in their indices). 
This means that the term involving $g_7$ vanishes identically even at the quantum level. Hence, $g_7$ does not enter in our equations of motion and thus we can drop this term from the expression of $G_{mnpq}$ given in \eqref{Group_theory_U}.

\vspace*{.06in}Next, we consider the Lagrange multipliers. The first constraint identity $I_1$ is given by
\be
:N^{mn}J\lambda^\alpha (\gamma_m)_{\alpha\beta}(K_1)^\beta_n:-\f{1}{2}:JJ\lambda^\alpha (\gamma^n)_{\alpha\beta}(K_1)^\beta_n:-\alpha':J\p\lambda^\alpha\gamma^n_{\alpha\beta}(K_1)^\beta_n:\ =\ 0
\ee
Using the expression of $(K_1)^\beta_n$ given in \eqref{Lag_multi_ansatz}, we find that the last two terms in the left hand side of the above expression vanish identically and the equation reduces to 
\be
c_1 k^r:N^{mn}J\lambda^\alpha (\gamma_m)_{\alpha\beta}(\gamma_r)^{\beta\sigma}\Psi_{n\sigma}:\ =\ 0
\ee
Again following the similar steps as described after equation \eqref{vanishg7} and noting that $J$ and $N^{mn}$ have trivial OPE, we find that this equation is identically satisfied and hence $c_1$ does not enter into our equations of motion. Thus, we can drop $(K_1)^\beta_n$ from the equations given in the previous subsection. 

\vspace*{.06in}Finally, we consider the term involving $c_9$ in the Lagrange multiplier $(K_3)^{\alpha\beta}_n$. After contracting $(I_3)_{\alpha\beta}^n$ with the term involving $c_9$ of $(K_3)^{\alpha\beta}_n$, we find that the last two terms of the constraint identity $I_3$ vanish identically whereas the first term vanishes by using the similar argument as given below equation \eqref{vanishg7}. Thus, we can also drop the term involving $c_9$ from our equation of motions.  

\vspace*{.06in}We are now ready to solve the equations of motion and determine the unknown coefficients appearing in the superfields. 
\subsection{Solving for Unknown Coefficients}
To determine the unknown coefficients in superfields, we put \eqref{Group_theory_U} and \eqref{Lag_multi_ansatz} in the equations of motion given in subsection \ref{eqsetup} and analyze them one by one. Some of the equations will determine the unknown coefficients while others will be satisfied identically. The Mathematica package GAMMA is very helpful for doing these calculations \cite{Gamma}.

\vspace*{.06in}The first five equations\footnote{To extract the information from the 3rd equation, it is convenient to contract it with 1-form, 3-form and 5-forms. This gives rise to 3 different equations which determine $f_2,f_5, h_2$ and $h_3$. Similarly, $g_1$ and $h_1$ can be determined from the 5th equation by contracting it with the 3-form.} of the previous subsection give\footnote{In general, some of the coefficients appearing in the superfields are determined by more than one equations. But, their values always agree. This also shows the consistency of the equations with our ansatz.}
\be
&&\hspace*{-.5in}f_1=- \f{18}{\alpha'} \quad,\quad f_2=\f{288i}{\alpha'}\quad,\quad f_3=\f{36i}{\alpha'}\quad,\quad  f_4=\f{12}{(\alpha')^2}\quad,\quad f_5\ =\ -\f{4i}{\alpha'}\non\\[.4cm]
&&\hspace*{-.5in}h_1=\f{2}{\alpha'}\quad,\qquad  h_2=-\f{576i}{\alpha'}\quad,\qquad h_3=-\f{144i}{\alpha'}\quad,\qquad   a=-\f{1}{(\alpha')^2}\non\\[.4cm]
&&\hspace*{-.5in}g_1\ =\ -\f{432}{\alpha'}\label{coeffiecient5}
\ee

Next, we contract the combined 6th and 7th equation \eqref{combined6th7th} with $\gamma^p_{\alpha\beta}$ and $\gamma^{pqr}_{\alpha\beta}$ and use \eqref{Group_theory_U} to find
\be
g_2=\f{48}{(\alpha')^2}\qquad,\quad\qquad g_3=-\f{192}{\alpha'}
\ee
Multiplying with a 5-form $\gamma_{\alpha\beta}^{pqrst}$ does not give any new information due to \eqref{ddcons345}. 

\vspace*{.06in}The equation 8 gives 
\be
f_4=\f{4}{(\alpha')^2}-8a\non
\ee
which is identically satisfied by the values of $f_4$ and $a$ given in equation \eqref{coeffiecient5}. Next, using \eqref{coeffiecient5}, the 9th equation determines
\be
c_{13}=\f{24}{\alpha'}\quad,\qquad c_{14}=-\f{24}{\alpha'}\quad,\qquad c_{15}=-30 \quad,\qquad c_{16}=192\label{c13-c16}
\ee

The equation 10 gives 
\be
10c_{13}+2c_{14}-\f{1}{\alpha'}c_{16}=0\non
\ee
Similarly, equation 11 gives
\be
&&\hspace*{-.3in}-\f{if_2\alpha'}{2}+-11i\alpha'f_3-21(\alpha')^2f_4
+\f{g_1\alpha'}{2} -2(\alpha')^2g_2+\f{\alpha'g_3}{4}\non\\
&&\hspace*{.5in}=\;\; -72-\alpha'\left(10c_{13}+2c_{14}-\f{1}{\alpha'}c_{16}\right)
\non
\ee
Both of these equations are identically satisfied by \eqref{coeffiecient5} and \eqref{c13-c16}. 

\vspace*{.06in}Next, the 12th equation gives
\be
c_{17}=\f{63i}{16}\quad,\qquad c_{18}=\f{3}{8\alpha'}\quad,\qquad c_{19}=-\f{9}{16\alpha'}\quad,\qquad c_{20}=-\f{57}{16}\label{c17-c20}
\ee
Using \eqref{c17-c20} and \eqref{coeffiecient5}, the equations resulting from 13th and 14th equations, namely, 
\be
c_{18}+7c_{19}-\f{c_{20}}{\alpha'}=0\non
\ee
and
\be
a(\alpha')^2-\f{f_4(\alpha')^2}{8}
+\f{if_5\alpha}{2}+\f{i\alpha'}{32}\left(\f{h_2}{3}-h_3\right)-h_1\alpha'=-1+\alpha' \left(c_{18}+7c_{19}-\f{c_{20}}{\alpha'}\right)
\non
\ee
are identically satisfied. 

\vspace*{.06in}Further, the equations 15, 16, 17 and 18 are identically satisfied by the ansatz in \eqref{Group_theory_U} and \eqref{Lag_multi_ansatz} without putting any restriction on the coefficients. 

\vspace*{.06in}Next, the 19th equation on using \eqref{g4=g5} gives 
\be
g_4\ =\ g_5\ =\ \f{4i}{(\alpha')^2}\label{g4=g5det}
\ee
Similarly, on dropping the terms involving $g_7$ and $c_9$ as discussed in the previous subsection and using equation \eqref{g4=g5det}, the 20th equation gives 
\be
g_6\ =\ -\f{12}{\alpha'}\quad,\quad c_{10}\ =\ \f{i}{\alpha'}\quad,\quad c_{11}\ =\ 0\quad,\quad c_{12}\ =\ -\f{i}{6\alpha'}\label{c10-c12}
\ee

Next, using equations \eqref{coeffiecient5}, \eqref{g4=g5det} and \eqref{c10-c12}, the equations resulting from 21st and 22nd equations, namely
\be
c_{10}-c_{11}+6c_{12}=0
\ee
and,
\be
&&ia(\alpha')^2+\f{i\alpha'}{48}\left(g_2\alpha'+\f{g_3}{2}\right)
-\f{f_5\alpha'}{2}+\f{(\alpha')^2}{16}(g_4+g_5)
=\f{i}{2}-\alpha'\Bigl(c_{10}-c_{11}+6c_{12}\Bigl)
\non
\ee
are identically satisfied. 

\vspace*{.06in}Finally, the 23rd, 24th equations along with \eqref{asym26and5} and \eqref{sym26} determine the Lagrange multiplier superfields $(K_2)^\beta_{mnp}$ and $(K_6)^\beta_m$. On dropping the terms involving $g_7$ and the Lagrange multiplier $(K_1)^\beta_n$ from these equations as discussed in the previous subsection and using the other coefficients determined so far, these 4 equations give
\be
c_2\ =\ -\f{96i}{5\alpha'}\quad,\quad c_3\ =\ -\f{72i}{5\alpha'}\quad,\quad c_4\ =\ \f{72i}{5\alpha'}\quad,\quad c_5\ =\ \f{8i}{5\alpha'}\non\\[.3cm]
c_6\ =\ -\f{8i}{5\alpha'}\quad,\quad c_7\ =\ 96i\quad,\quad c_8\ =\ \f{24i}{5\alpha'}\quad,\quad c_{21}\ =\ -9i
\ee
We have now determined all the coefficients appearing in the ansatz for superfields and the Lagrange multipliers. We have also exhausted all the equations of motion. With these coefficients, the BRST equation of motion $QU = \p_{\mathbb{R}} V$ is now identically satisfied. This establishes that our ansatz for various superfields with the values of coefficients determined in this section indeed gives the correct integrated vertex for the first massive states. The final expression of the integrated vertex operator $U$ including the numerical coefficients in the ansatz is given in equations \eqref{general_Ures} and \eqref{Group_theory_Ures}.

\sectiono{Conclusion} \label{discus}
We have constructed the integrated form of the first massive vertex operator of open strings in the pure spinor formalism. Since the vertex operator is solely expressed in terms of the superfields $B_{mnp}, G_{mn}$ and $\Psi_{m\alpha}$, using the theta expansion results given in \cite{theta_exp}, one can readily obtain the theta expansion of the integrated vertex in terms of only the physical fields $b_{mnp}, g_{mn}$ and $\psi_{m\alpha}$. This, therefore, demonstrates that the integrated vertex operator thus constructed is in terms of the physical degrees of freedom only.  

\vspace*{0.06in} This construction can also be used to obtain the first massive integrated vertex operator in the Heterotic string. For this, one simply need to take the tensor product of the vertex operator constructed here with the anti-holomorphic integrated vertex of the bosonic string. However some normalisation factors need to be accounted for while going to closed superstrings from open superstrings.

\vspace*{.06in}Previously, with only the unintegrated form of the massive vertex being known, the possible scattering amplitudes involving massless and first massive states that could be explicitly computed, were severely restricted. Knowing the integrated vertex now enables one to compute any amplitude upto two loop order involving arbitrary number of the massless and first massive states in the pure spinor formalism\footnote{It is the understanding of the authors that at present there are no unanimous consensus on computing full multiloop amplitudes in pure spinor formalism. But, also see\cite{BerkovitsNekrasov, BerkovitsJapan, Grassi2 }.}. The results of some amplitude calculations involving massive states will be reported in the future \cite{code_paper}.

\vspace*{.06in}The pure spinor constraints as well as the OPEs imply that the basic worldsheet operators satisfy non-linear constraints. This fact leads to several subtleties. In particular, it implies non-trivial identities which a subset of all worldsheet operators at a given conformal weight and ghost number will satisfy. We showed how to take into account all such constraints systematically in section \ref{Constraint_iden3}. This line of reasoning was based on its successful role in determining the unintegrated vertex \cite{Berkovits1} and is now further strengthened by the successful construction of the integrated form of the vertex. These evidences therefore suggest that we have indeed adopted the correct way of incorporating the effect of all such constraints at higher mass levels.

\vspace*{.06in}The general strategy outlined in section \textbf{\ref{strategy}} and the method given in appendix \ref{argument_gen_super}  for writing the ansatz do not explicitly or implicitly depend on the conformal weight and ghost number for which we eventually employed it. It is also to be noted that an identical strategy can be applied to construct even unintegrated vertex for any massive state, the only difference being the equation that one now needs to solve is $QV=0$. This leads us to conjecture that our strategy is very general and can be successfully implemented to determine integrated as well as unintegrated form of  vertex operators for all higher massive states in pure spinor formalism. We plan to explicitly test this in future works \cite{future1}.

\bigskip

\noindent{\bf Acknowledgments:}  
We are deeply thankful to Ashoke Sen for suggesting to look into the problem, for numerous illuminating discussions throughout the course of this work and for very insightful comments on the draft. We are also deeply thankful to Nathan Berkovits and Renann Lipinski Jusinskas for useful email exchanges and Kris Thielemans for the mathematica package OPEDefs. We also thank Thales Azevedo, Anirban Basu, Abhishek Chowdhury, Rajesh Gopakumar, Dileep P. Jatkar, R. Loganayagam, Anshuman Maharana, Swapnamay Mondal, Satchitananda Naik and Oliver Schlotterer for discussions. We also thank Nathan Berkovits and Rajesh Gopakumar for the comments on an earlier version of this draft. This research was supported in part by the Infosys Fellowship for the senior students. MV is also thankful to University of California, Los Angeles and University of California, Davis for the hospitality while this work was in progress. We also thank the people and Government of India for their continuous support for theoretical physics.

\appendix
\section{Motivating the ansatz} 
\label{motive}
\subsection{The polynomial dependance of vertex operators on the pure spinor ghost field}
\label{polynomial_dependence}
In writing the most general form of the integrated vertex in equation \eqref{general_U}, we assumed that it does not depend upon the $\bar\lambda\lambda$ factors. In this appendix, we justify this assumption. First we recall that the integrated vertex $U$ can also be determined by integrating the 
$b$ ghost around the unintegrated vertex $V$, i.e., 
\be
U(z) = \oint\f{dw}{2\pi i}\ b(w)V(z)
\ee
In the pure spinor formalism, the $b$ ghost is a composite operator which involves different powers of $\bar\lambda\lambda$ in the denominator \cite{Berkovits7}.
So, naively, one might expect that the integrated vertex will also involve different powers of  $\bar\lambda\lambda$ in denominator. However, it is possible to work 
in a gauge in which the vertex operators are independent of the $\bar\lambda\lambda$ terms. To see this, we recall from the RNS formalism that the
massive states also appear in the OPEs of the massless vertex operators. This allows us, in principle,  to construct the massive vertex operators from the
knowledge of the massless vertex operators. More specifically for open strings, this construction, pointed out to the authors by Nathan Berkovits, goes as follows. If $V_1, V_2$ are unintegrated and $U_1, U_2$ are integrated massless vertex operators respectively, then we have 
\be
Q U_1 = \p_{\mathbb{R}} V_1 \qquad\mbox{and}\qquad Q U_2 = \p_{\mathbb{R}} V_2\label{masslessmassless1}
\ee 
We now take the contour integral of $U_1$ around the integrand of $U_2$ and define 
\be
U_3(z) \equiv  \oint\f{dw}{2\pi i} U_1(w) U_2(z)\label{A.33333}
\ee
Acting on this with the BRST operator $Q$ and using \eqref{masslessmassless1}, we obtain
\be
QU_3= \oint\f{dw}{2\pi i} U_1(w) QU_2(z) = \oint\f{dw}{2\pi i} U_1(w) \p_zV_2(z) \equiv \p_z V_3 \label{QU_3}
\ee
where,
\be
V_3(z) \equiv  \oint\f{dw}{2\pi i} U_1(w) V_2(z)\label{A.55555}
\ee
and in the first equality in \eqref{QU_3}, we have used the fact that $\oint dw\ \p_{\mathbb{R}} V_1(w)$ is zero. 

\vspace*{.07in}Now, if we choose the momentum $k_1$ and $k_2$ of $U_1$ and $U_2$  to satisfy  
\be
(k_1+k_2)^2=2 k_1\cdot k_2 \equiv (k_3)^2=-m^2 =-\f{n}{\alpha'}
\ee
then, by construction, the $V_3$ and $U_3$ will be unintegrated and integrated massive vertex operators respectively of open string states at mass level $n$. 

\vspace*{.07in}{One might ask how do we know that the $U_3$ and $V_3$ as defined in \eqref{A.33333} and \eqref{A.55555} do not vanish. To answer this question, we recall that the OPE of two massless vertex operators necessarily contain the massive vertex operators (this is necessary for the consistency of the theory and is well known from the RNS formalism). Now, the integrated vertices $U_1$ and $U_2$ have conformal weight one. Hence, by dimensional analysis, it is easy to see that the integrand involving the integrated massive vertex operator can only appear at the first order pole in \eqref{A.33333} and hence its contour integral can't vanish. By a similar argument, we see that $V_3$ as defined in \eqref{A.55555} can't vanish.}

\vspace*{.07in}Since the massless vertices can be chosen to be independent of $\bar \lambda\lambda$ in denominator \cite{Berkovits}, this construction shows 
that the massive vertices can also be constructed without using the $\bar \lambda \lambda$ in the denominator. Moreover, since the massless vertices do not involve
$JJ$ and $\p J$ terms, the above construction also shows why $JJ$ and $\p J$ terms do not appear in the massive vertices. In appendix \ref{argument_gen_super}, we
 give another argument for this based on group theory.

\subsection{General form of the Superfields}
\label{argument_gen_super}
In this appendix, we give the method for writing down the ansatz \eqref{Group_theory_U} and \eqref{Lag_multi_ansatz} for the massive superfields which appear in the vertex operators and Lagrange multipliers. The same method can be very easily generalized for the construction of any massive vertex operator in the pure spinor formalism. 

\vspace*{.06in} We start by arguing that the superfields appearing in the integrated vertex operator must be expressible in terms of the basic superfields
$B_{mnp}, G_{mn}$ and $\Psi_{m\alpha}$. This follows because as shown in \cite{Berkovits1, theta_exp}, the superfields appearing in the full set of superspace
equations of motion can be expressed solely in terms of any of the basic superfields $\Psi_{m\alpha}, B_{mnp}$ or $G_{mn}$. Thus, the vertex operators should
be expressible entirely in terms of any of these basic superfields\footnote{This is similar to the case of the massless vertices. The massless vertices are also
	expressed entirely in terms of the superfields which appear in the $\mathcal{N}=1$ super Yang Mills equations of motion in 10 dimensions.}. For the unintegrated 
vertex  operator \eqref{vertex_ansatz}, this can be seen using equations \eqref{2.14} and  \eqref{D_Gmn1}-\eqref{cons_theta=0}. From this, it is also clear that 
if we want to express the entire vertex operator in terms of only one or two basic superfield, we need to use the supercovariant derivative. However, if we use all
the 3 superfields, then we can avoid the use of supercovariant derivatives (since the supercovariant derivative of the basic superfields can be expressed in terms of
the basic superfields without supercovariant derivative using equations \eqref{D_Gmn1}-\eqref{cons_theta=0}). 

\vspace*{.07in} We shall make use of all the 3 basic superfields $B_{mnp}, G_{mn}$ and $\Psi_{m\alpha}$. Thus, due to equations \eqref{D_Gmn1}-\eqref{cons_theta=0}, 
the relation between the superfields in integrated vertex operator and these basic superfields can be expressed without using the super covariant derivative. 
Moreover, whatever be the functional form of these superfields, the Lorentz invariance implies that they can only involve 3 basic superfields, the momentum vector,
the space-time metric $\eta^{mn}$ and the gamma matrices. Thus, the functional dependence of all the superfields in the momentum space is
\be
\mbox{ Superfields in $U$}  = f(B_{mnp}, G_{mn}, \Psi_{m\alpha}, k_m, \eta^{mn},\mbox{Gamma Matrices})\non
\ee
Our goal now is to determine these functions. This can be done by making use of the group representation theory. To see this, we note that the physical degrees
of freedom (encoded in the fields $\psi_{m\alpha}, b_{mnp}$ and $g_{mn}$) should match on both sides at each order in the theta expansion of the above equation. 
Moreover, since the right hand side does not involve supercovariant derivative, it follows that we can equate the coefficients in the theta expansion of the 
superfield in the left hand side at a given order with the coefficient at the same order in the theta expansion of the right hand side\footnote{Note that if
	we have a superspace equation of the form $S_\alpha= D_\alpha T$, then the $\ell^{th}$ order component of the superfield $S_\alpha$ will be 
	related to $(\ell-1)^{th}$ and $(\ell+1)^{th}$ order components of the superfield $T$. However, if we have an equation of the form $S_\alpha=R_\alpha$,
	then the $\ell^{th}$ order component of $S_\alpha$ will be related to $\ell^{th}$ order component of $R_\alpha$.  }. Since the right hand side involve only the 
basic superfields $\Psi_{m\alpha}, B_{mnp}$ and $G_{mn}$, it follows that any given order theta component of the superfield in the left side is related to the
same order theta component of the basic superfields $\Psi_{m\alpha}, B_{mnp}$ and $G_{mn}$. We now focus on the theta independent component. Using above argument,
it follows that the theta independent components of the superfields in the left hand side must be expressible in terms of only the theta independent components of
the basic superfields $B_{mnp}, G_{mn}$ and $\Psi_{m\alpha}$, namely $b_{mnp}, g_{mn}$ and $\psi_{m\alpha}$\footnote{This is not true for the massless states. 
	One reason for this is that given a differential equation of the form $D_\alpha S= B_\alpha$ (where $S$ and $B_\alpha$ are some superfields encoding the 
	information about the massless states), one can't invert this to write an expression for $S$ in terms of some differential operator acting on $B_\alpha$ 
	since $k^2=0$ for the massless states. This is unlike the massive states where we can always invert this kind of equations.}. 

\vspace*{.07in} Thus, for the theta independent components of the superfields in the integrated vertex, our problem has reduced to finding the correct physical
degrees of freedom and to express them in terms of $b_{mnp}, g_{mn}$ and $\psi_{m\alpha}$. The covariant expression for the full superfield can then be obtained
by replacing $b_{mnp}, g_{mn}$ and $\psi_{m\alpha}$ by $B_{mnp}, G_{mn}$ and $\Psi_{m\alpha}$ respectively. The validity of this procedure can be justified by the
fact that it gives an operator $U$ which satisfies the correct BRST equation $QU=\p_{\mathbb{R}} V$. Once we have an operator $U$ which satisfies this equation, we are guaranteed
that it is the correct integrated vertex irrespective of how we arrive at it.

\vspace*{.06in} Now, the correct degrees of freedom in the theta independent components of the superfields can be obtained by looking at their index structure
and using the group theory. In the rest frame, the physical fields $b_{mnp}, g_{mn}$ and $\psi_{m\alpha}$ form the \textbf{84, 44} and \textbf{128} representations
of the little group SO(9). Thus, to determine the correct physical degrees of freedom in the theta independent components of the superfields, we need to find the
number of \textbf{84, 44} and \textbf{128} representations of SO(9) in their theta independent components in the rest frame.  

\vspace*{.07in} We shall illustrate this method by some examples now. First, consider the superfield $C_{mn}$ in \eqref{general_U}. Since, it is anti symmetric
in its indices $m$ and $n$, its only non zero components in the rest frame can be $C_{0a}$ and $C_{ab}$. These can only form \textbf{9} and \textbf{36}
representation of SO(9) and hence can't contain the physical massive fields. Thus, $C_{mn}$ must be zero. Similarly, all the superfields whose theta independent
components cannot form the \textbf{84, 44} or \textbf{128} representations of SO(9), must be zero. Next, we consider the superfield $G^\beta_{mn}$. Since, it is also
anti symmetric in its vector indices $m$ and $n$, going to the rest frame, we find that its non zero components can only be $ G^\beta_{0a}$ and $G^\beta_{ab}$. Our
goal is to look for representations of SO(9) corresponding to the physical states. Now, the index structure of $G^\beta_{0a}$ implies that its theta independent 
component forms the product representation $\textbf{16}\times \textbf{9}$ which contains one $\textbf{128}$. Similarly, $G^\beta_{ab}$ contains one $\textbf{128}$. 
This means that the theta independent component of $G^\beta_{mn}$ should contain two representations of $\textbf{128}$ and hence there should be two terms involving
$\Psi_{m\alpha}$ in the expansion of $G^\beta_{mn}$ in terms of the basic fields $B_{mnp}, G_{mn}$ and $\Psi_{m\alpha}$.  After finding the correct number of terms,
the next step is to write down the form of $G^\beta_{mn}$ so that it has two terms involving $\Psi_{m\alpha}$. Taking into account the on shell conditions 
\eqref{cons_theta=0}, we find
\be
G^{ \beta}_{pq}\ =\ g_2\gamma_{[p}^{ \beta\sigma}\Psi_{q]\sigma}+g_3 k^r\gamma_r^{ \beta\sigma}k_{[p}\Psi_{q]\sigma}
\ee
where $g_2$ and $g_3$ are some unknown coefficients which need to be determined.

\color{black}

\end{document}